\documentclass[aps,prl,preprint,superscriptaddress]{revtex4-1}
\usepackage{amsmath}
\usepackage{amssymb}
\usepackage{gensymb}
\usepackage{graphicx}
\usepackage{epstopdf}
\usepackage{float}
\begin{document}
	
	\title{Spatially resolved x-ray studies of liquid crystals with strongly developed bond-orientational order}
		
	\author{I.~A.~Zaluzhnyy}
	\affiliation{Deutsches Elektronen-Synchrotron DESY, Notkestra{\ss}e 85, D-22607 Hamburg, Germany}
	\affiliation{National Research Nuclear University MEPhI (Moscow Engineering Physics Institute), Kashirskoe shosse 31, 115409 Moscow, Russia}
	
	\author{R.~P.~Kurta}
	\altaffiliation{Present address: European XFEL GmbH, Albert-Einstein-Ring 19, D-22761 Hamburg, Germany}
	\affiliation{Deutsches Elektronen-Synchrotron DESY, Notkestra{\ss}e 85, D-22607 Hamburg, Germany}
	
	\author{E.~A.~Sulyanova}
	\affiliation{Institute of Crystallography, Russian Academy of Sciences, Leninskii prospect 59, 119333 Moscow, Russia}
		
	\author{O.~Y.~Gorobtsov}
	\affiliation{Deutsches Elektronen-Synchrotron DESY, Notkestra{\ss}e 85, D-22607 Hamburg, Germany}
	\affiliation{NRC Kurchatov Institute, Akademika Kurchatova pl. 1, 123182 Moscow, Russia}
		
	\author{A.~G.~Shabalin}
	\affiliation{Deutsches Elektronen-Synchrotron DESY, Notkestra{\ss}e 85, D-22607 Hamburg, Germany}
	
	\author{A.~V.~Zozulya}
	\affiliation{Deutsches Elektronen-Synchrotron DESY, Notkestra{\ss}e 85, D-22607 Hamburg, Germany}
	
	\author{A.~P.~Menushenkov}
	\affiliation{National Research Nuclear University MEPhI (Moscow Engineering Physics Institute), Kashirskoe shosse 31, 115409 Moscow, Russia}
	
	\author{M.~Sprung}
	\affiliation{Deutsches Elektronen-Synchrotron DESY, Notkestra{\ss}e 85, D-22607 Hamburg, Germany}
	
	\author{B.~I.~Ostrovskii}
	\email[Corresponding authors: ]{ostenator@gmail.com and ivan.vartaniants@desy.de}
	\affiliation{Institute of Crystallography, Russian Academy of Sciences, Leninskii prospect 59, 119333 Moscow, Russia}
	
	\author{I.~A.~Vartanyants}
	\email[Corresponding authors: ]{ostenator@gmail.com and ivan.vartaniants@desy.de}
	\affiliation{Deutsches Elektronen-Synchrotron DESY, Notkestra{\ss}e 85, D-22607 Hamburg, Germany}
	\affiliation{National Research Nuclear University MEPhI (Moscow Engineering Physics Institute), Kashirskoe shosse 31, 115409 Moscow, Russia}

	\date{\today}
	
	\begin{abstract}
We present an x-ray study of freely suspended hexatic films of the liquid crystal 3(10)OBC.
Our results reveal spatial inhomogeneities of the bond-orientational (BO) order in the vicinity of the hexatic-smectic phase transition and the formation of large scale hexatic domains at lower temperatures.
Deep in the hexatic phase up to 25 successive sixfold BO order parameters have been directly determined by means of angular x-ray cross-correlation analysis (XCCA).
Such strongly developed hexatic order allowed us to determine higher order correction terms in the scaling relation predicted by the multicritical scaling theory over a full temperature range of the hexatic phase existence.
	\end{abstract}
	
	\maketitle

The influence of angular correlations on structural and physical properties of complex fluids, colloidal suspensions and liquid crystals (LCs) remains one of fundamental and unresolved problems in modern condensed matter physics \cite{Chaikin}.
A prominent example of a system with angular correlations is the hexatic phase that combines the properties of both crystals and liquids \cite{Strandburg1992}.
The two-dimensional (2D) hexatic phase shows a sixfold quasi-long range bond-orientational (BO) order, while the positional order is short range \cite{Nelson2002}.
The hexatic phase is a general phenomenon that was observed in a number of systems of various physical nature, such as 2D colloids \cite{Murray1987, Kusner1994,Keim2007}, electrons at the surface of helium \cite{Glattli1988}, 2D superconducting vortexes \cite{Murray1990, Guillamon2009} and, particularly, in liquid crystals \cite{Pindak1981,Chou1997,Brock1986,Stoebe1995}.

The hexatic phase was predicted by Halperin and Nelson \cite{Halperin1978} as an intermediate state in 2D crystal melting.
According to their theory the hexatic phase arises as a consequence of the broken translational symmetry of a 2D crystal induced by dissociation of dislocation pairs.
This mechanism does not work in 3D crystals, however, the 3D hexatic phase was observed experimentally in LCs \cite{Pindak1981}.
The multicritical scaling theory (MCST) developed by Aharony and coworkers \cite{Aharony1986} based on renormalization group approach to critical phenomena enabled quantitative characterization of the BO order in the hexatic phase and, particularly, allowed to study a crossover from 2D to 3D behavior \cite{Chou1997,Stoebe1992}.
In spite of the extensive experimental and theoretical work the origin of the hexatic phase in LCs and the features of the hexatic - smectic phase transition remain puzzling and controversial.

The structure of hexatics is traditionally studied by means of x-ray or electron diffraction in a single-domain area of a hexatic film (see for reviews \cite{Jeu2003,Brock1989,Brock1989a}).
The quantitative characteristics of the BO order, the so-called BO order parameters \cite{Brock1986}, are typically determined by fitting the measured azimuthal intensity distribution by the Fourier cosine series.
In contrast to this approach in the present work we performed spatially resolved x-ray diffraction studies of free standing LC films.
Measured x-ray data were analyzed by means of direct Fourier transformation and by using angular x-ray cross-correlation analysis (XCCA) \cite{Wochner2009,Altarelli2010,Kurta2012,Kurta2013a}.
The latter method enabled a direct determination of sixfold BO order parameters from the ensemble of diffraction patterns without applying a fitting procedure \cite{Kurta2013}.
	
The coherent x-ray scattering experiment on smectic LC membranes was performed at the beamline P10 of the PETRA III facility at DESY in Hamburg.
The incident photon energy was 13 keV with the flux $3\cdot10^{10}$~photons/sec.
The detector (Pilatus 1M) was positioned in transmission geometry at the distance of 263 mm from the sample.
The beam at the sample plane was focused to 2$\times$3~$\mu$m$^2$ (vertical vs. horizontal) at full width at half maximum (FWHM) by a set of compound refractive lenses \cite{Zozulya2012}.
A specially designed sample stage was used to prepare LC films \textit{in~situ} and maintain the temperature with an accuracy of 0.005~$\degree$C (see for experimental details \cite{Kurta2013}).
	
In our experiment we have used the LC compound 3(10)OBC (n-propyl-4-n-decyloxy-biphenyl-4-carboxylate)  \cite{Stoebe1995,Huang1989}.
The hexatic-smectic phase transition was found at $T \approx 66.3 \pm 0.1~\degree$C and the material crystallizes below T$\approx$54~$\degree$C.
The films of 3(10)OBC sample were drawn across a small circular glass aperture of 2 mm in diameter inside the chamber at 10~$\degree$C above the temperature of the hexatic-smectic phase transition.
The thickness of the films was determined to be in the range of $5-7~\mu$m.
Films were slowly cooled down to observe the formation and development of the hexatic phase.
For spatially resolved studies at each temperature the sample was scanned in the plane perpendicular to the incident beam direction in the region of 100$\times$100~$\mu$m$^2$ with 11~$\mu$m step size.
Exposure time of 0.6 seconds  was chosen to sustain the nondestructive regime of measurements.
	
Typical diffraction patterns in the hexatic and smectic phases corrected for background scattering and horizontal synchrotron polarization are shown in Figs. 1(a-c).
At high temperatures in the smectic phase the diffraction pattern has a form of a broad uniform ring [Fig. 1(a)] due to the absence of angular correlations and short-range positional order.
During the hexatic-smectic phase transition this scattering ring splits into six arcs [Fig. 1(b)], revealing the sixfold rotational symmetry of the in-plane molecular arrangement.
While the temperature decreases all six arcs become narrower both in azimuthal and radial directions, indicating the simultaneous development of the BO and positional order [Fig. 1(c)].

	\begin{figure}
	\includegraphics[width=\linewidth]{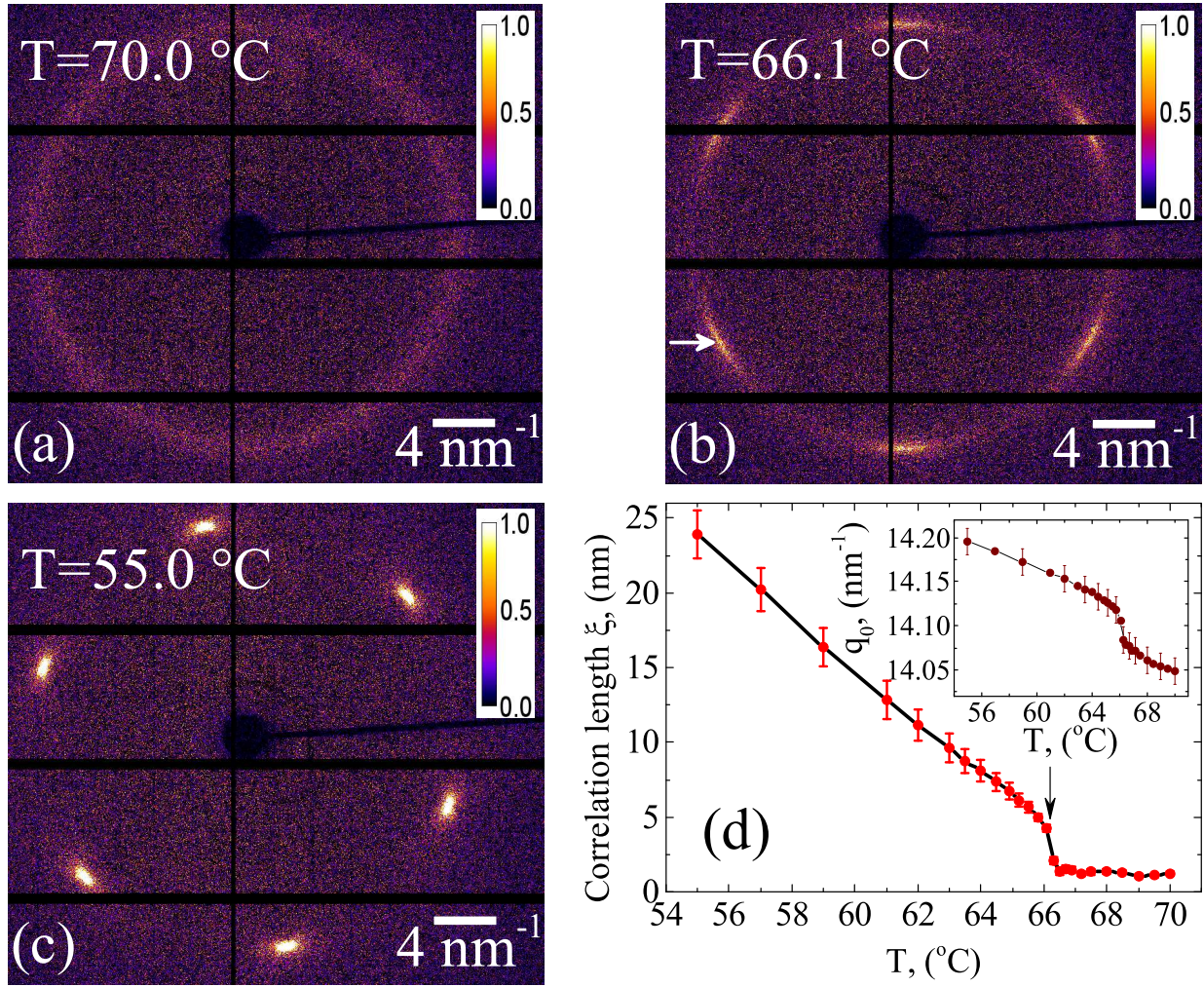}
	\caption{(a-c) Diffraction patterns from a LC film in the smectic (a) and hexatic phase (b,c). (d) Temperature dependence of the positional correlation length $\xi$ determined from the radial profile of the scattering peak marked by an arrow in (b). The temperature of the hexatic-smectic phase transition is specified by an arrow. In the inset the temperature dependence of the peak maximum position $q_0$ is shown.}
	\label{Fig1}
	\end{figure}

The in-plane structure of smectic films is characterized by the positional correlation length $\xi$, which is commonly defined as $\xi=1/\Delta q$, where $\Delta q$ is the half width at half maximum (HWHM) of a radial scan through the center of a diffraction peak \cite{Jeu2003}.
The value of $\xi$ determined by fitting the radial intensity profiles with the Lorentzian function [see Supplemental Materials] at each position in the sample and averaged over a set of $N=100$ diffraction patterns at each temperature  is shown in Fig. \ref{Fig1}(d).
In the smectic phase the positional correlation length $\xi$ was determined to be about 1.5 nm.
Close to the hexatic phase transition point at T$=$66.3~$\degree$C the correlation length starts to increase and upon further cooling down it reaches the value of 24 nm in the hexatic phase.
The simultaneous rise of the peak's maximum position $q_0$ and the presence of the inflection point on the $q_0$ temperature dependence [see inset in Fig. 1(d)] indicates the growing fluctuations of the BO order parameter in the vicinity of the hexatic-smectic phase transition \cite{Aeppli1984,Davey1984}.
	
To determine the details of the BO order formation and especially its spatial distribution in the vicinity of the hexatic-smectic phase transition region we performed an expansion of the scattered intensity measured at each position in the sample into a Fourier cosine series
$I(q,\phi)=I_0(q)+2\sum_{n=1}^{n=+\infty}|I_n(q)|\cos\left(n\phi-\phi_n(q)\right)$, where $|I_n(q)|exp(i\phi_n(q))=(1 / 2\pi)\int_{0}^{2\pi}I(q,\!\phi)exp(-in\phi) \mathrm{d}\phi$.
In this Fourier decomposition $(q, \phi)$ are the polar coordinates in the detector plane, $I_0(q)$ is the scattered intensity averaged over a scattering ring of a radius $q$, and $|I_n(q)|$ and $\phi_n(q)$ are the magnitude and phase of the $n-$th Fourier component (FC).
	
The in-plane orientation of the molecular bonds was determined at each spatial position using the phase $\phi_6(q_0)$ and magnitude $|I_6(q_0)|$ of the FC of the dominant 6-th order [see Supplemental Materials], calculated at the scattering peak maximum position $q=q_0$.
In the smectic phase $|I_6(q_0)|$ was found to be below the noise threshold due to the absence of angular correlations.
As soon as the temperature was decreased to the phase transition region the situation has dramatically changed (see Fig. 2(a,b)).
Remarkably, the 2D map shown in Fig. 2(a) reveals inhomogeneity in the spatial distribution of the hexatic phase at the temperature T$=$66.3~$\degree$C. This is clearly reflected in variation of the magnitude $|I_6(q_0)|$ across the sample.
Areas with larger magnitudes of the vectors correspond to the regions with higher degree of the BO order.
While cooling down the sample to the temperature T$=$66.1~$\degree$C a single domain with identical orientation of molecular bonds over the entire region of scanning was formed [Fig. 2(b)].
Such uniform BO order with the same orientation of molecular bonds persists upon cooling down into the crystalline phase.

	\begin{figure}
	\includegraphics[width=\linewidth]{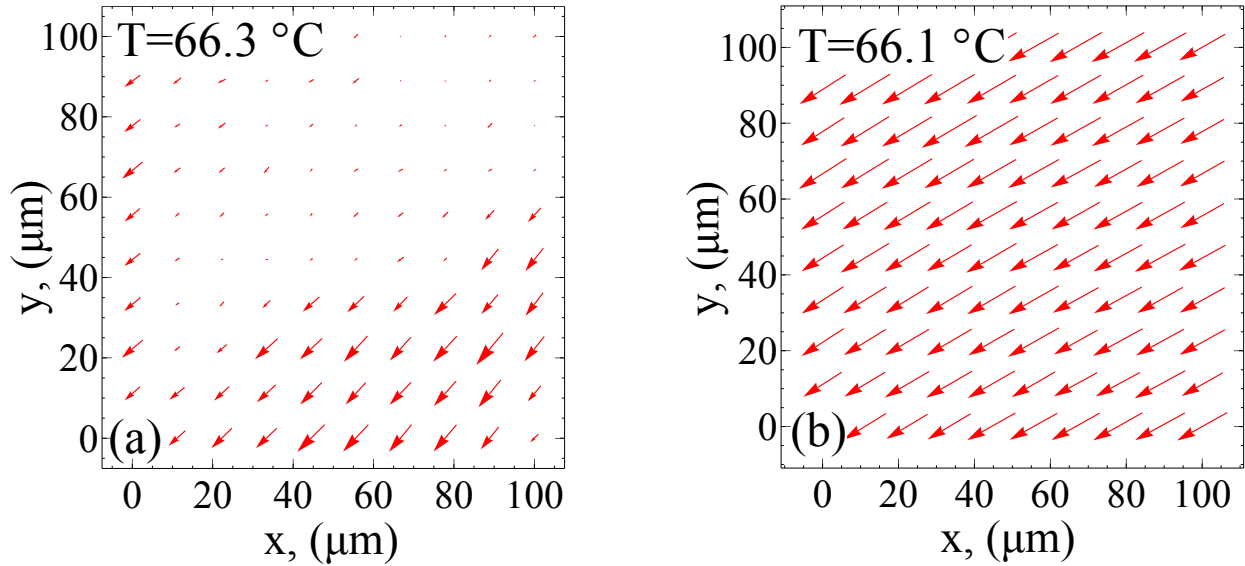}
	\caption{\label{Fig2} Spatially resolved 2D maps of the LC film structure at T$=$66.3~$\degree$C (a) and T$=$66.1~$\degree$C (b). Each vector in the plot corresponds to a certain position in the sample. Direction of a vector is associated with the angular position of the center of the peak marked with an arrow in Fig. 1(b), and the length of a vector is proportional to the magnitude $|I_6(q_0)|$. The length of the vectors in (a) is magnified five times as compared to the vectors in (b).}
	\end{figure}

Next, we applied XCCA \cite{Altarelli2010} to determine variation of the BO order parameters in the hexatic phase as a function of temperature.
Such approach is not sensitive to small spatial variations of the molecular bonds orientation and gives more reliable information on BO order parameters than fitting of azimuthal intensity distribution.
The basic element of this approach is the two-point cross-correlation function (CCF) \cite{Clark1983,Wochner2009,Altarelli2010}
	\begin{equation}
	\label{Eq2}
	G(q,\Delta)=\langle I(q,\phi)I(q,\phi+\Delta)\rangle_{\phi} \ ,
	\end{equation}
where $\langle\ldots\rangle_{\phi}$ is the angular average over a ring of radius $q$ and $0\le\Delta\!<\!2\pi$ is the angular coordinate.
Since $G(q,\Delta)$ is a real even function, it can be decomposed into the Fourier cosine series, $G(q,\Delta)=G_0(q,\Delta)+2\sum_{n=1}^{n=+\infty}G_n(q)\cos(n\Delta)$,
where $G_n(q)$ are the FCs of the CCF.
In the present case of a single domain the FCs $G_n(q)$ averaged over a set of $N$ diffraction patterns are related to the FCs of intensity $I_n(q)$ by a simple relation $\langle G_n(q)\rangle_N=|I_n(q)|^2$ \cite{Kurta2013a}.
	
\begin{figure}
		\includegraphics[width=\linewidth]{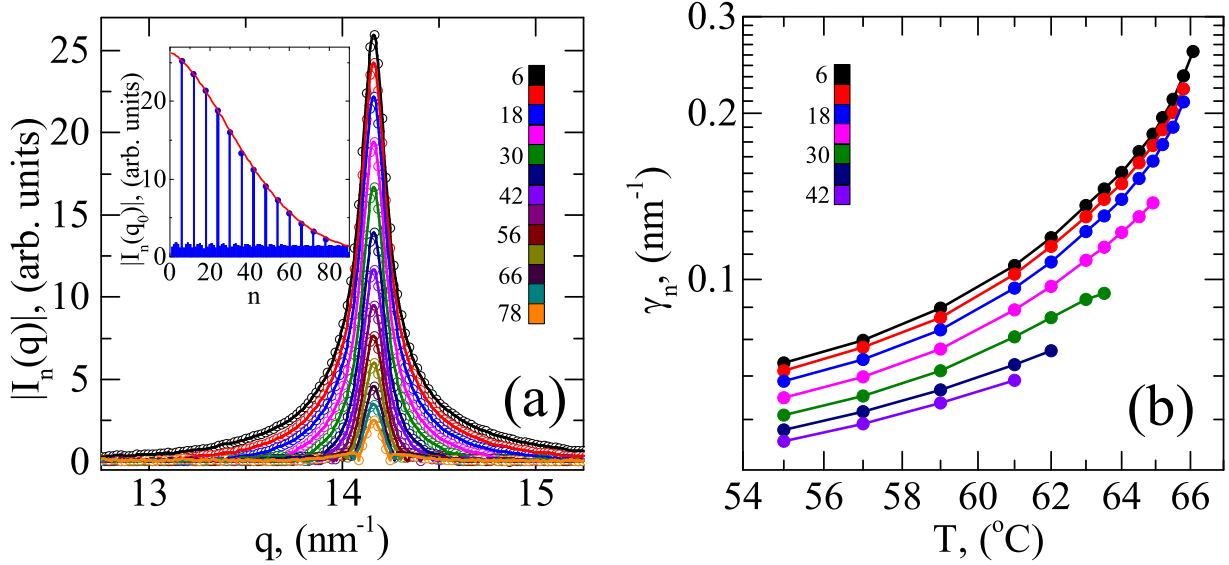}
		\caption{ (a) Magnitudes of the FCs of intensity $|I_n (q)|$ with $n=6,12,\ldots,78$ as a function of $q$ at the temperature T$=$61.$\degree$C.
Solid lines are SRL and Lorentzian fits to the
experimental data (points) (see text for details).
(inset) Magnitudes of the FCs $|I_n (q_0)|$ at $q_0$=14.16~nm$^{-1}$ as a function of the order $n$.
Red line represents the fitting using the MCST.
(b) Temperature dependence of the HWHM $\gamma_n$ of the FCs of intensity.}
\label{Fig3}
\end{figure}

The FCs of intensity $|I_n(q)| = \sqrt{\langle G_n(q)\rangle_N}$, where $N=100$ in our case, as a function of $q$ measured at the temperature T$=$61.0~$\degree$C are presented in Fig. \ref{Fig3}(a).
As one can see, we determined an unprecedented number of harmonics in the hexatic phase of the 3(10)OBC film.
To characterize the BO order parameters we used the maximum values of the magnitudes of the FCs $|I_n (q_0)|$ ($q_0$=14.16~nm$^{-1}$) that are shown in the inset of Fig. \ref{Fig3}(a). %for  $1\le n \le 78$.
One can clearly see that the magnitudes of the FCs systematically decrease as a function of the harmonic number $n$, in agreement with MCST \cite{Aharony1986} (see red line).
In the ideal hexatic phase only FCs $|I_n(q_0)|$ of the orders $n=6m$, where $m=1,2,3\ldots$, contribute to the Fourier decomposition \cite{Brock1986}.
A finite nonzero contribution of other FCs appear due to small background contribution and other uncompensated experimental factors.
A threshold was defined as the average value of these components, and only FCs with the magnitudes above this threshold were considered in the further analysis.

Our analysis has shown that it is not possible to characterize the FCs of intensity as a function of $q$ presented in Fig. \ref{Fig3}(a) by single functional form \cite{Harmonics_q}.
In our experiment the FCs of lower orders $n$ were better described by the square root Lorentzian (SRL) and the higher orders by the Lorentzian functions, respectively [see Supplemental Materials].
Results of the fitting of the few first FCs as a function of temperature are presented in Fig. \ref{Fig3}(b).
We can observe a gradual decrease of the HWHM of the peaks $\gamma_n$ with the harmonic number $n$ and upon cooling \cite{Kurta2013}.

\begin{figure}
		\includegraphics[width=\linewidth]{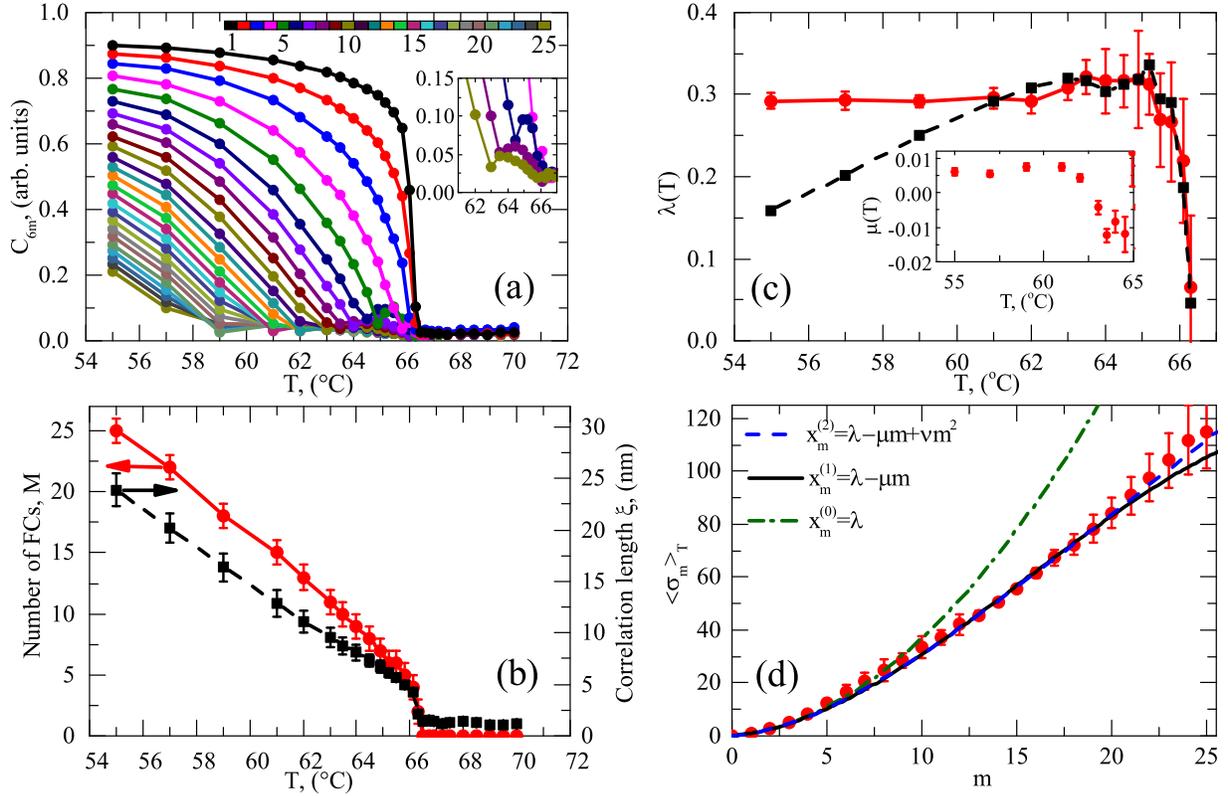}
		\caption{
(a) Temperature dependence of the normalized BO order parameters $C_{6m}$.
In the inset the non-monotonic behavior of the BO order parameters of the orders $m = 4,..., 10$ is shown.
(b) Temperature dependence of the total number of FCs $M$ and positional correlation length $\xi$ [see Fig. \ref{Fig1}(d)].
(c) Temperature dependence of the parameter $\lambda (T)$ (red points) in the scaling relation (\ref{Eq4M}) from the MCST.
Compare with the temperature dependence of the same parameter (black squares) when the linear correction term in the scaling relation (\ref{Eq4M}) is neglected.
(d) Temperature averaged values of $\langle \sigma_m \rangle_T$ (red circles) and their fit with the scaling relation (\ref{Eq4M}) (black line).
Results of fitting without the correction term (green dash-dot line) and with the 2-nd order correction term (blue dash line) are also shown.
\label{Fig4}
}
	\end{figure}

The evolution of the BO order as a function of temperature can be described by the set of independent normalized BO order parameters, $C_{6m}=|(I_{6m} (q_0)) / (I_0 (q_0))|$  defined in \cite{Brock1986}.
The number of nonzero coefficients $C_{6m}$ and their magnitude characterizes a degree of the BO order development.
The temperature dependence of the BO order parameters $C_{6m}$ is presented in Fig. \ref{Fig4}(a).
In the smectic phase all parameters $C_{6m}$ have values below the threshold level.
The FCs of the 6-th and 12-th order appear at the phase transition point T$=$66.3~$\degree$C, and further components exceed the threshold level one after another while cooling down below the phase transition temperature.
Finally, we observed unusually large number $m$=25 successive BO order parameters at the temperature T$=$55.0~$\degree$C.
Interestingly, we also observed a subtle non-monotonic behavior of the FCs of the orders $m = 4, ..., 10$ in the temperature range $63~\degree$C$<T<66~\degree$C [see inset in Fig. \ref{Fig4}(a)].
The temperature dependence of the $C_{6m}$ parameters confirms a general trend described earlier for 2D \cite{Chou1997} and 3D \cite{Brock1986,Aharony1986,Brock1989a} hexatics. However, the maximum number of the registered BO order parameters in these experiments was significantly smaller, typically less than ten.
The total number of the determined FCs, $M$, as a function of temperature is shown in Fig \ref{Fig4}(b).
The correlation between this number and the positional correlation length $\xi$ directly indicates strong coupling between the BO and positional order in the hexatic phase of 3(10)OBC film.

A quantitative comparison of the BO parameters of different orders can be made on the basis of the MCST \cite{Aharony1986}.
This theory predicts the following scaling relation for the BO order parameters, $C_{6m}=(C_6 )^{\sigma_m}$, with the exponent $\sigma_m$ of the form
\begin{equation}
    \sigma_m=m+x_m\cdot m \cdot (m-1) , \
    x^{(1)}_m=\lambda (T)-\mu (T)m \ ,
\label{Eq4M}
\end{equation}
where the parameters $\lambda(T)\cong0.3$ and $\mu(T)\cong8\cdot10^{-3}$ are given by the MCST for the 3D hexatic phase \cite{Brock1989,Brock1989a}.
At each temperature the experimentally determined BO order parameters $C_{6m}$ were fitted by the scaling relation (\ref{Eq4M}) (an example of such fit 
for the temperature $T = 61.0~\degree$C is shown in the inset in Fig. \ref{Fig3}(a)).
The temperature dependence of the determined parameters $\lambda(T)$ and $\mu(T)$ is presented in Fig. \ref{Fig4}(c) by red circles [see Supplemental Materials].
As we can see, the scaling relation (\ref{Eq4M}) accurately describes our experimental data in the entire temperature range of the hexatic phase existence.
At low temperatures the parameters of the scaling law reach practically the constant values, $\lambda(T) \approx 0.29 \pm 0.01$ and $\mu(T) \approx 0.007 \pm 0.001$, that are in excellent agreement with the theoretical predictions of the MCST for 3D hexatics \cite{Aharony1986}.
In the previous experiments \cite{Brock1986,Aharony1986} only the parameter $\lambda$ has been determined.
Here, due to the presence of the numerous harmonics, we were able to deduce also the value of the first order correction term $\mu$.
Our results also show that if the first order correction term in $x^{(1)}_m$ is neglected (shown by black squares in Fig. \ref{Fig4}(c)), then we get a strong deviation from the theoretically predicted value $\lambda(T)=0.3$ at low temperatures, where many FCs are present.
We also found that in the vicinity of the hexatic-smectic phase transition the parameters $\lambda(T)$ and $\mu(T)$ rapidly decrease in a narrow temperature range of $\Delta T \approx 1~\degree$C [Fig. \ref{Fig4}(c)] (similar observation was made for the parameter $\lambda(T)$ in \cite{Aharony1986}).

Since the exponents $\sigma_m$ in the scaling relation (\ref{Eq4M}) appear to be almost independent of temperature, one can consider the temperature averaged values $\langle \sigma_m \rangle_T$ to determine the parameters $\lambda$ and $\mu$  by applying equations (\ref{Eq4M}) \cite{Aharony1986}.
The experimentally obtained values of $\langle\sigma_m\rangle_T$ are shown in Fig. \ref{Fig4}(d).
It is well seen that the term $x^{(1)}_m$ with  $\lambda=0.31 \pm 0.015$ and $\mu=0.009 \pm 0.001$ determined by fitting $\langle \sigma_m \rangle_T$ data points with equations (\ref{Eq4M}), describes very well all orders $m$ measured in the experiment \cite{NOTE1}.
At the same time if the first order correction term in $x^{(1)}_m$ is neglected then the scaling relation (\ref{Eq4M}) correctly describes the behavior of the parameter $\langle \sigma_m \rangle_T$ only up to the 7-th order but fails to characterize accurately higher orders $m$.
We also observed that at high values of $m \geq 20$ there is a small but constant deviation of the measured and fitted values.
Statistical analysis based on the F-test \cite{Bevington2003} revealed that the scaling relation term in the form $x^{(2)}_m = \lambda - \mu m + \nu m^2$
is statistically significant with the significance level of 0.02.
The fitting of the experimentally determined values of $\langle \sigma_m \rangle_T$ with this scaling relation [shown by dash blue line in Fig. \ref{Fig4}(d)] gave for the parameter $\nu$ a small value of $\nu\approx 1\cdot10^{-4}$. Determining this second order correction value theoretically using renormalization theory would require to perform $\epsilon$-expansion of the order higher than $\epsilon^2$ ($\epsilon=4-d$, where $d$ is the space dimension) \cite{Aharony1986}. 
	
In summary, we investigated the in-plane structure of free standing hexatic films of 3(10)OBC LC by means of x-ray diffraction technique.
A micron-sized spatially resolved measurements allowed to construct 2D maps of the in-plane molecular bonds orientations that revealed inhomogeneity of the BO order at the hexatic-smectic phase transition region.
An exceptional high number ($m=25$) of the successive BO order parameters have been directly determined at low temperatures in the hexatic phase by means of XCCA.
Our results support the validity of the MCST in the whole temperature range of existence of the 3D hexatic phase.
We have confirmed experimentally the value of the the first order correction term in the scaling relation of the BO order parameters that was predicted earlier by the MCST. 
Moreover, we also revealed a small second order correction term that was not yet determined theoretically.
We expect that our work will stimulate both theoretical studies and development of experimental techniques for quantitative investigation of the effects of long scale angular correlations on the structure and physical properties of diverse LC phases and complex fluids.

	\begin{acknowledgments}
	 We acknowledge support of this project and discussions with E. Weckert. We are thankful to C.~C. Huang, W.~H. de Jeu, E.~I. Kats, V.~V. Lebedev, A.~R. Muratov, and V.~M. Kaganer for fruitful discussions.
	This work was partially supported by the Virtual Institute VH-VI-403 of the Helmholtz Association.
	The work of I.A.Z. and B.I.O. was partially supported by the Russian Science Foundation (grant 14-12-00475).
	\end{acknowledgments}

	\bibliography{Ref}

\begin{thebibliography}{32}%
\makeatletter
\providecommand \@ifxundefined [1]{%
 \@ifx{#1\undefined}
}%
\providecommand \@ifnum [1]{%
 \ifnum #1\expandafter \@firstoftwo
 \else \expandafter \@secondoftwo
 \fi
}%
\providecommand \@ifx [1]{%
 \ifx #1\expandafter \@firstoftwo
 \else \expandafter \@secondoftwo
 \fi
}%
\providecommand \natexlab [1]{#1}%
\providecommand \enquote  [1]{``#1''}%
\providecommand \bibnamefont  [1]{#1}%
\providecommand \bibfnamefont [1]{#1}%
\providecommand \citenamefont [1]{#1}%
\providecommand \href@noop [0]{\@secondoftwo}%
\providecommand \href [0]{\begingroup \@sanitize@url \@href}%
\providecommand \@href[1]{\@@startlink{#1}\@@href}%
\providecommand \@@href[1]{\endgroup#1\@@endlink}%
\providecommand \@sanitize@url [0]{\catcode `\\12\catcode `\$12\catcode
  `\&12\catcode `\#12\catcode `\^12\catcode `\_12\catcode `\%12\relax}%
\providecommand \@@startlink[1]{}%
\providecommand \@@endlink[0]{}%
\providecommand \url  [0]{\begingroup\@sanitize@url \@url }%
\providecommand \@url [1]{\endgroup\@href {#1}{\urlprefix }}%
\providecommand \urlprefix  [0]{URL }%
\providecommand \Eprint [0]{\href }%
\providecommand \doibase [0]{http://dx.doi.org/}%
\providecommand \selectlanguage [0]{\@gobble}%
\providecommand \bibinfo  [0]{\@secondoftwo}%
\providecommand \bibfield  [0]{\@secondoftwo}%
\providecommand \translation [1]{[#1]}%
\providecommand \BibitemOpen [0]{}%
\providecommand \bibitemStop [0]{}%
\providecommand \bibitemNoStop [0]{.\EOS\space}%
\providecommand \EOS [0]{\spacefactor3000\relax}%
\providecommand \BibitemShut  [1]{\csname bibitem#1\endcsname}%
\let\auto@bib@innerbib\@empty
%</preamble>
\bibitem [{\citenamefont {Chaikin}\ and\ \citenamefont
  {Lubensky}(1995)}]{Chaikin}%
  \BibitemOpen
  \bibfield  {author} {\bibinfo {author} {\bibfnamefont {P.}~\bibnamefont
  {Chaikin}}\ and\ \bibinfo {author} {\bibfnamefont {T.}~\bibnamefont
  {Lubensky}},\ }\href@noop {} {\emph {\bibinfo {title} {Principles of
  Condensed Matter Physics}}}\ (\bibinfo  {publisher} {Cambridge University
  Press},\ \bibinfo {year} {1995})\BibitemShut {NoStop}%
\bibitem [{\citenamefont {Strandburg}(1992)}]{Strandburg1992}%
  \BibitemOpen
  \bibfield  {author} {\bibinfo {author} {\bibfnamefont {K.~J.}\ \bibnamefont
  {Strandburg}},\ }\href
  {http://www.springer.com/materials/book/978-1-4612-7680-7} {\emph {\bibinfo
  {title} {Bond-Orirntational Order in Condensed Matter Systems}}},\ edited by\
  \bibinfo {editor} {\bibfnamefont {K.~J.}\ \bibnamefont {Strandburg}}\
  (\bibinfo  {publisher} {Springer-Verlag New York, Inc},\ \bibinfo {year}
  {1992})\BibitemShut {NoStop}%
\bibitem [{\citenamefont {Nelson}(2002)}]{Nelson2002}%
  \BibitemOpen
  \bibfield  {author} {\bibinfo {author} {\bibfnamefont {D.~R.}\ \bibnamefont
  {Nelson}},\ }\href@noop {} {\emph {\bibinfo {title} {Defects and Geometry in
  Condensed Matter Physics}}}\ (\bibinfo  {publisher} {Cambridge University
  Press},\ \bibinfo {year} {2002})\BibitemShut {NoStop}%
\bibitem [{\citenamefont {Murray}\ and\ \citenamefont
  {Van~Winkle}(1987)}]{Murray1987}%
  \BibitemOpen
  \bibfield  {author} {\bibinfo {author} {\bibfnamefont {C.~A.}\ \bibnamefont
  {Murray}}\ and\ \bibinfo {author} {\bibfnamefont {D.~H.}\ \bibnamefont
  {Van~Winkle}},\ }\href {\doibase 10.1103/PhysRevLett.58.1200} {\bibfield
  {journal} {\bibinfo  {journal} {Phys. Rev. Lett.}\ }\textbf {\bibinfo
  {volume} {58}},\ \bibinfo {pages} {1200} (\bibinfo {year}
  {1987})}\BibitemShut {NoStop}%
\bibitem [{\citenamefont {Kusner}\ \emph {et~al.}(1994)\citenamefont {Kusner},
  \citenamefont {Mann}, \citenamefont {Kerins},\ and\ \citenamefont
  {Dahm}}]{Kusner1994}%
  \BibitemOpen
  \bibfield  {author} {\bibinfo {author} {\bibfnamefont {R.~E.}\ \bibnamefont
  {Kusner}}, \bibinfo {author} {\bibfnamefont {J.~A.}\ \bibnamefont {Mann}},
  \bibinfo {author} {\bibfnamefont {J.}~\bibnamefont {Kerins}}, \ and\ \bibinfo
  {author} {\bibfnamefont {A.~J.}\ \bibnamefont {Dahm}},\ }\href {\doibase
  10.1103/PhysRevLett.73.3113} {\bibfield  {journal} {\bibinfo  {journal}
  {Phys. Rev. Lett.}\ }\textbf {\bibinfo {volume} {73}},\ \bibinfo {pages}
  {3113} (\bibinfo {year} {1994})}\BibitemShut {NoStop}%
\bibitem [{\citenamefont {Keim}\ \emph {et~al.}(2007)\citenamefont {Keim},
  \citenamefont {Maret},\ and\ \citenamefont {von Gr\"unberg}}]{Keim2007}%
  \BibitemOpen
  \bibfield  {author} {\bibinfo {author} {\bibfnamefont {P.}~\bibnamefont
  {Keim}}, \bibinfo {author} {\bibfnamefont {G.}~\bibnamefont {Maret}}, \ and\
  \bibinfo {author} {\bibfnamefont {H.~H.}\ \bibnamefont {von Gr\"unberg}},\
  }\href {\doibase 10.1103/PhysRevE.75.031402} {\bibfield  {journal} {\bibinfo
  {journal} {Phys. Rev. E}\ }\textbf {\bibinfo {volume} {75}},\ \bibinfo
  {pages} {031402} (\bibinfo {year} {2007})}\BibitemShut {NoStop}%
\bibitem [{\citenamefont {Glattli}\ \emph {et~al.}(1988)\citenamefont
  {Glattli}, \citenamefont {Andrei},\ and\ \citenamefont
  {Williams}}]{Glattli1988}%
  \BibitemOpen
  \bibfield  {author} {\bibinfo {author} {\bibfnamefont {D.~C.}\ \bibnamefont
  {Glattli}}, \bibinfo {author} {\bibfnamefont {E.~Y.}\ \bibnamefont {Andrei}},
  \ and\ \bibinfo {author} {\bibfnamefont {F.~I.~B.}\ \bibnamefont
  {Williams}},\ }\href {\doibase 10.1103/PhysRevLett.60.420} {\bibfield
  {journal} {\bibinfo  {journal} {Phys. Rev. Lett.}\ }\textbf {\bibinfo
  {volume} {60}},\ \bibinfo {pages} {420} (\bibinfo {year} {1988})}\BibitemShut
  {NoStop}%
\bibitem [{\citenamefont {Murray}\ \emph {et~al.}(1990)\citenamefont {Murray},
  \citenamefont {Gammel}, \citenamefont {Bishop}, \citenamefont {Mitzi},\ and\
  \citenamefont {Kapitulnik}}]{Murray1990}%
  \BibitemOpen
  \bibfield  {author} {\bibinfo {author} {\bibfnamefont {C.~A.}\ \bibnamefont
  {Murray}}, \bibinfo {author} {\bibfnamefont {P.~L.}\ \bibnamefont {Gammel}},
  \bibinfo {author} {\bibfnamefont {D.~J.}\ \bibnamefont {Bishop}}, \bibinfo
  {author} {\bibfnamefont {D.~B.}\ \bibnamefont {Mitzi}}, \ and\ \bibinfo
  {author} {\bibfnamefont {A.}~\bibnamefont {Kapitulnik}},\ }\href {\doibase
  10.1103/PhysRevLett.64.2312} {\bibfield  {journal} {\bibinfo  {journal}
  {Phys. Rev. Lett.}\ }\textbf {\bibinfo {volume} {64}},\ \bibinfo {pages}
  {2312} (\bibinfo {year} {1990})}\BibitemShut {NoStop}%
\bibitem [{\citenamefont {Guillamon}\ \emph {et~al.}(2009)\citenamefont
  {Guillamon}, \citenamefont {Suderow}, \citenamefont {Fernandez-Pacheco},
  \citenamefont {Sese}, \citenamefont {Cordoba}, \citenamefont {De~Teresa},
  \citenamefont {Ibarra},\ and\ \citenamefont {Vieira}}]{Guillamon2009}%
  \BibitemOpen
  \bibfield  {author} {\bibinfo {author} {\bibfnamefont {I.}~\bibnamefont
  {Guillamon}}, \bibinfo {author} {\bibfnamefont {H.}~\bibnamefont {Suderow}},
  \bibinfo {author} {\bibfnamefont {A.}~\bibnamefont {Fernandez-Pacheco}},
  \bibinfo {author} {\bibfnamefont {J.}~\bibnamefont {Sese}}, \bibinfo {author}
  {\bibfnamefont {R.}~\bibnamefont {Cordoba}}, \bibinfo {author} {\bibfnamefont
  {J.~M.}\ \bibnamefont {De~Teresa}}, \bibinfo {author} {\bibfnamefont {M.~R.}\
  \bibnamefont {Ibarra}}, \ and\ \bibinfo {author} {\bibfnamefont
  {S.}~\bibnamefont {Vieira}},\ }\href {http://dx.doi.org/10.1038/nphys1368}
  {\bibfield  {journal} {\bibinfo  {journal} {Nat. Phys.}\ }\textbf {\bibinfo
  {volume} {5}},\ \bibinfo {pages} {651} (\bibinfo {year} {2009})}\BibitemShut
  {NoStop}%
\bibitem [{\citenamefont {Pindak}\ \emph {et~al.}(1981)\citenamefont {Pindak},
  \citenamefont {Moncton}, \citenamefont {Davey},\ and\ \citenamefont
  {Goodby}}]{Pindak1981}%
  \BibitemOpen
  \bibfield  {author} {\bibinfo {author} {\bibfnamefont {R.}~\bibnamefont
  {Pindak}}, \bibinfo {author} {\bibfnamefont {D.~E.}\ \bibnamefont {Moncton}},
  \bibinfo {author} {\bibfnamefont {S.~C.}\ \bibnamefont {Davey}}, \ and\
  \bibinfo {author} {\bibfnamefont {J.~W.}\ \bibnamefont {Goodby}},\ }\href
  {\doibase 10.1103/PhysRevLett.46.1135} {\bibfield  {journal} {\bibinfo
  {journal} {Phys. Rev. Lett.}\ }\textbf {\bibinfo {volume} {46}},\ \bibinfo
  {pages} {1135} (\bibinfo {year} {1981})}\BibitemShut {NoStop}%
\bibitem [{\citenamefont {Chou}\ \emph {et~al.}(1997)\citenamefont {Chou},
  \citenamefont {Ho},\ and\ \citenamefont {Hui}}]{Chou1997}%
  \BibitemOpen
  \bibfield  {author} {\bibinfo {author} {\bibfnamefont {C.-F.}\ \bibnamefont
  {Chou}}, \bibinfo {author} {\bibfnamefont {J.~T.}\ \bibnamefont {Ho}}, \ and\
  \bibinfo {author} {\bibfnamefont {S.~W.}\ \bibnamefont {Hui}},\ }\href
  {\doibase 10.1103/PhysRevE.56.592} {\bibfield  {journal} {\bibinfo  {journal}
  {Phys. Rev. E}\ }\textbf {\bibinfo {volume} {56}},\ \bibinfo {pages} {592}
  (\bibinfo {year} {1997})}\BibitemShut {NoStop}%
\bibitem [{\citenamefont {Brock}\ \emph {et~al.}(1986)\citenamefont {Brock},
  \citenamefont {Aharony}, \citenamefont {Birgeneau}, \citenamefont
  {Evans-Lutterodt}, \citenamefont {Litster}, \citenamefont {Horn},
  \citenamefont {Stephenson},\ and\ \citenamefont {Tajbakhsh}}]{Brock1986}%
  \BibitemOpen
  \bibfield  {author} {\bibinfo {author} {\bibfnamefont {J.~D.}\ \bibnamefont
  {Brock}}, \bibinfo {author} {\bibfnamefont {A.}~\bibnamefont {Aharony}},
  \bibinfo {author} {\bibfnamefont {R.~J.}\ \bibnamefont {Birgeneau}}, \bibinfo
  {author} {\bibfnamefont {K.~W.}\ \bibnamefont {Evans-Lutterodt}}, \bibinfo
  {author} {\bibfnamefont {J.~D.}\ \bibnamefont {Litster}}, \bibinfo {author}
  {\bibfnamefont {P.~M.}\ \bibnamefont {Horn}}, \bibinfo {author}
  {\bibfnamefont {G.~B.}\ \bibnamefont {Stephenson}}, \ and\ \bibinfo {author}
  {\bibfnamefont {A.~R.}\ \bibnamefont {Tajbakhsh}},\ }\href {\doibase
  10.1103/PhysRevLett.57.98} {\bibfield  {journal} {\bibinfo  {journal} {Phys.
  Rev. Lett.}\ }\textbf {\bibinfo {volume} {57}},\ \bibinfo {pages} {98}
  (\bibinfo {year} {1986})}\BibitemShut {NoStop}%
\bibitem [{\citenamefont {Stoebe}\ \emph {et~al.}(1995)\citenamefont {Stoebe},
  \citenamefont {Jin}, \citenamefont {Mach},\ and\ \citenamefont
  {Huang}}]{Stoebe1995}%
  \BibitemOpen
  \bibfield  {author} {\bibinfo {author} {\bibfnamefont {T.}~\bibnamefont
  {Stoebe}, \bibfnamefont {T.E}}, \bibinfo {author} {\bibfnamefont
  {A.}~\bibnamefont {Jin}}, \bibinfo {author} {\bibfnamefont {P.}~\bibnamefont
  {Mach}}, \ and\ \bibinfo {author} {\bibfnamefont {C.}~\bibnamefont {Huang}},\
  }\href {http://dx.doi.org/10.1007/BF01438955} {\bibfield  {journal} {\bibinfo
   {journal} {Int. J. Thermophys.}\ }\textbf {\bibinfo {volume} {16}},\
  \bibinfo {pages} {33} (\bibinfo {year} {1995})}\BibitemShut {NoStop}%
\bibitem [{\citenamefont {Halperin}\ and\ \citenamefont
  {Nelson}(1978)}]{Halperin1978}%
  \BibitemOpen
  \bibfield  {author} {\bibinfo {author} {\bibfnamefont {B.~I.}\ \bibnamefont
  {Halperin}}\ and\ \bibinfo {author} {\bibfnamefont {D.~R.}\ \bibnamefont
  {Nelson}},\ }\href {http://link.aps.org/doi/10.1103/PhysRevLett.41.121}
  {\bibfield  {journal} {\bibinfo  {journal} {Phys. Rev. Lett.}\ }\textbf
  {\bibinfo {volume} {41}},\ \bibinfo {pages} {121} (\bibinfo {year}
  {1978})}\BibitemShut {NoStop}%
\bibitem [{\citenamefont {Aharony}\ \emph {et~al.}(1986)\citenamefont
  {Aharony}, \citenamefont {Birgeneau}, \citenamefont {Brock},\ and\
  \citenamefont {Litster}}]{Aharony1986}%
  \BibitemOpen
  \bibfield  {author} {\bibinfo {author} {\bibfnamefont {A.}~\bibnamefont
  {Aharony}}, \bibinfo {author} {\bibfnamefont {R.~J.}\ \bibnamefont
  {Birgeneau}}, \bibinfo {author} {\bibfnamefont {J.~D.}\ \bibnamefont
  {Brock}}, \ and\ \bibinfo {author} {\bibfnamefont {J.~D.}\ \bibnamefont
  {Litster}},\ }\href {\doibase 10.1103/PhysRevLett.57.1012} {\bibfield
  {journal} {\bibinfo  {journal} {Phys. Rev. Lett.}\ }\textbf {\bibinfo
  {volume} {57}},\ \bibinfo {pages} {1012} (\bibinfo {year}
  {1986})}\BibitemShut {NoStop}%
\bibitem [{\citenamefont {Stoebe}\ \emph {et~al.}(1992)\citenamefont {Stoebe},
  \citenamefont {Geer}, \citenamefont {Huang},\ and\ \citenamefont
  {Goodby}}]{Stoebe1992}%
  \BibitemOpen
  \bibfield  {author} {\bibinfo {author} {\bibfnamefont {T.}~\bibnamefont
  {Stoebe}}, \bibinfo {author} {\bibfnamefont {R.}~\bibnamefont {Geer}},
  \bibinfo {author} {\bibfnamefont {C.~C.}\ \bibnamefont {Huang}}, \ and\
  \bibinfo {author} {\bibfnamefont {J.~W.}\ \bibnamefont {Goodby}},\ }\href
  {\doibase 10.1103/PhysRevLett.69.2090} {\bibfield  {journal} {\bibinfo
  {journal} {Phys. Rev. Lett.}\ }\textbf {\bibinfo {volume} {69}},\ \bibinfo
  {pages} {2090} (\bibinfo {year} {1992})}\BibitemShut {NoStop}%
\bibitem [{\citenamefont {de~Jeu}\ \emph {et~al.}(2003)\citenamefont {de~Jeu},
  \citenamefont {Ostrovskii},\ and\ \citenamefont {Shalaginov}}]{Jeu2003}%
  \BibitemOpen
  \bibfield  {author} {\bibinfo {author} {\bibfnamefont {W.~H.}\ \bibnamefont
  {de~Jeu}}, \bibinfo {author} {\bibfnamefont {B.~I.}\ \bibnamefont
  {Ostrovskii}}, \ and\ \bibinfo {author} {\bibfnamefont {A.~N.}\ \bibnamefont
  {Shalaginov}},\ }\href {\doibase 10.1103/RevModPhys.75.181} {\bibfield
  {journal} {\bibinfo  {journal} {Rev. Mod. Phys.}\ }\textbf {\bibinfo {volume}
  {75}},\ \bibinfo {pages} {181} (\bibinfo {year} {2003})}\BibitemShut
  {NoStop}%
\bibitem [{\citenamefont {Brock}\ \emph
  {et~al.}(1989{\natexlab{a}})\citenamefont {Brock}, \citenamefont {Noh},
  \citenamefont {McClain}, \citenamefont {Litster}, \citenamefont {Birgeneau},
  \citenamefont {Aharony}, \citenamefont {Horn},\ and\ \citenamefont
  {Liang}}]{Brock1989}%
  \BibitemOpen
  \bibfield  {author} {\bibinfo {author} {\bibfnamefont {J.}~\bibnamefont
  {Brock}}, \bibinfo {author} {\bibfnamefont {D.}~\bibnamefont {Noh}}, \bibinfo
  {author} {\bibfnamefont {B.}~\bibnamefont {McClain}}, \bibinfo {author}
  {\bibfnamefont {J.}~\bibnamefont {Litster}}, \bibinfo {author} {\bibfnamefont
  {R.}~\bibnamefont {Birgeneau}}, \bibinfo {author} {\bibfnamefont
  {A.}~\bibnamefont {Aharony}}, \bibinfo {author} {\bibfnamefont
  {P.}~\bibnamefont {Horn}}, \ and\ \bibinfo {author} {\bibfnamefont
  {J.}~\bibnamefont {Liang}},\ }\href {http://dx.doi.org/10.1007/BF01307386}
  {\bibfield  {journal} {\bibinfo  {journal} {Z. Phys. B - Condensed Matter}\
  }\textbf {\bibinfo {volume} {74}},\ \bibinfo {pages} {197} (\bibinfo {year}
  {1989}{\natexlab{a}})}\BibitemShut {NoStop}%
\bibitem [{\citenamefont {Brock}\ \emph
  {et~al.}(1989{\natexlab{b}})\citenamefont {Brock}, \citenamefont {Birgeneau},
  \citenamefont {Litster},\ and\ \citenamefont {Aharony}}]{Brock1989a}%
  \BibitemOpen
  \bibfield  {author} {\bibinfo {author} {\bibfnamefont {J.~D.}\ \bibnamefont
  {Brock}}, \bibinfo {author} {\bibfnamefont {R.~J.}\ \bibnamefont
  {Birgeneau}}, \bibinfo {author} {\bibfnamefont {D.}~\bibnamefont {Litster}},
  \ and\ \bibinfo {author} {\bibfnamefont {A.}~\bibnamefont {Aharony}},\ }\href
  {\doibase 10.1080/00107518908213772} {\bibfield  {journal} {\bibinfo
  {journal} {Contemporary Physics}\ }\textbf {\bibinfo {volume} {30}},\
  \bibinfo {pages} {321} (\bibinfo {year} {1989}{\natexlab{b}})}\BibitemShut
  {NoStop}%
\bibitem [{\citenamefont {Wochner}\ \emph {et~al.}(2009)\citenamefont
  {Wochner}, \citenamefont {Gutt}, \citenamefont {Autenrieth}, \citenamefont
  {Demmer}, \citenamefont {Bugaev}, \citenamefont {Ortiz}, \citenamefont
  {Duri}, \citenamefont {Zontone}, \citenamefont {Gr\"{u}bel},\ and\
  \citenamefont {Dosch}}]{Wochner2009}%
  \BibitemOpen
  \bibfield  {author} {\bibinfo {author} {\bibfnamefont {P.}~\bibnamefont
  {Wochner}}, \bibinfo {author} {\bibfnamefont {C.}~\bibnamefont {Gutt}},
  \bibinfo {author} {\bibfnamefont {T.}~\bibnamefont {Autenrieth}}, \bibinfo
  {author} {\bibfnamefont {T.}~\bibnamefont {Demmer}}, \bibinfo {author}
  {\bibfnamefont {V.}~\bibnamefont {Bugaev}}, \bibinfo {author} {\bibfnamefont
  {A.~D.}\ \bibnamefont {Ortiz}}, \bibinfo {author} {\bibfnamefont
  {A.}~\bibnamefont {Duri}}, \bibinfo {author} {\bibfnamefont {F.}~\bibnamefont
  {Zontone}}, \bibinfo {author} {\bibfnamefont {G.}~\bibnamefont {Gr\"{u}bel}},
  \ and\ \bibinfo {author} {\bibfnamefont {H.}~\bibnamefont {Dosch}},\ }\href
  {\doibase 10.1073/pnas.0905337106} {\bibfield  {journal} {\bibinfo  {journal}
  {PNAS}\ }\textbf {\bibinfo {volume} {106}},\ \bibinfo {pages} {11511}
  (\bibinfo {year} {2009})}\BibitemShut {NoStop}%
\bibitem [{\citenamefont {Altarelli}\ \emph {et~al.}(2010)\citenamefont
  {Altarelli}, \citenamefont {Kurta},\ and\ \citenamefont
  {Vartanyants}}]{Altarelli2010}%
  \BibitemOpen
  \bibfield  {author} {\bibinfo {author} {\bibfnamefont {M.}~\bibnamefont
  {Altarelli}}, \bibinfo {author} {\bibfnamefont {R.~P.}\ \bibnamefont
  {Kurta}}, \ and\ \bibinfo {author} {\bibfnamefont {I.~A.}\ \bibnamefont
  {Vartanyants}},\ }\href {\doibase 10.1103/PhysRevB.82.104207} {\bibfield
  {journal} {\bibinfo  {journal} {Phys. Rev. B}\ }\textbf {\bibinfo {volume}
  {82}},\ \bibinfo {pages} {104207} (\bibinfo {year} {2010})}\BibitemShut
  {NoStop}%
\bibitem [{\citenamefont {Kurta}\ \emph {et~al.}(2012)\citenamefont {Kurta},
  \citenamefont {Altarelli}, \citenamefont {Weckert},\ and\ \citenamefont
  {Vartanyants}}]{Kurta2012}%
  \BibitemOpen
  \bibfield  {author} {\bibinfo {author} {\bibfnamefont {R.~P.}\ \bibnamefont
  {Kurta}}, \bibinfo {author} {\bibfnamefont {M.}~\bibnamefont {Altarelli}},
  \bibinfo {author} {\bibfnamefont {E.}~\bibnamefont {Weckert}}, \ and\
  \bibinfo {author} {\bibfnamefont {I.~A.}\ \bibnamefont {Vartanyants}},\
  }\href {\doibase 10.1103/PhysRevB.85.184204} {\bibfield  {journal} {\bibinfo
  {journal} {Phys. Rev. B}\ }\textbf {\bibinfo {volume} {85}},\ \bibinfo
  {pages} {184204} (\bibinfo {year} {2012})}\BibitemShut {NoStop}%
\bibitem [{\citenamefont {Kurta}\ \emph
  {et~al.}(2013{\natexlab{a}})\citenamefont {Kurta}, \citenamefont
  {Altarelli},\ and\ \citenamefont {Vartanyants}}]{Kurta2013a}%
  \BibitemOpen
  \bibfield  {author} {\bibinfo {author} {\bibfnamefont {R.~P.}\ \bibnamefont
  {Kurta}}, \bibinfo {author} {\bibfnamefont {M.}~\bibnamefont {Altarelli}}, \
  and\ \bibinfo {author} {\bibfnamefont {I.~A.}\ \bibnamefont {Vartanyants}},\
  }\href {\doibase 10.1155/2013/959835} {\bibfield  {journal} {\bibinfo
  {journal} {Adv. Condens. Matter Phys.}\ }\textbf {\bibinfo {volume} {2013}},\
  \bibinfo {pages} {959835} (\bibinfo {year} {2013}{\natexlab{a}})}\BibitemShut
  {NoStop}%
\bibitem [{\citenamefont {Kurta}\ \emph
  {et~al.}(2013{\natexlab{b}})\citenamefont {Kurta}, \citenamefont
  {Ostrovskii}, \citenamefont {Singer}, \citenamefont {Gorobtsov},
  \citenamefont {Shabalin}, \citenamefont {Dzhigaev}, \citenamefont {Yefanov},
  \citenamefont {Zozulya}, \citenamefont {Sprung},\ and\ \citenamefont
  {Vartanyants}}]{Kurta2013}%
  \BibitemOpen
  \bibfield  {author} {\bibinfo {author} {\bibfnamefont {R.~P.}\ \bibnamefont
  {Kurta}}, \bibinfo {author} {\bibfnamefont {B.~I.}\ \bibnamefont
  {Ostrovskii}}, \bibinfo {author} {\bibfnamefont {A.}~\bibnamefont {Singer}},
  \bibinfo {author} {\bibfnamefont {O.~Y.}\ \bibnamefont {Gorobtsov}}, \bibinfo
  {author} {\bibfnamefont {A.}~\bibnamefont {Shabalin}}, \bibinfo {author}
  {\bibfnamefont {D.}~\bibnamefont {Dzhigaev}}, \bibinfo {author}
  {\bibfnamefont {O.~M.}\ \bibnamefont {Yefanov}}, \bibinfo {author}
  {\bibfnamefont {A.~V.}\ \bibnamefont {Zozulya}}, \bibinfo {author}
  {\bibfnamefont {M.}~\bibnamefont {Sprung}}, \ and\ \bibinfo {author}
  {\bibfnamefont {I.~A.}\ \bibnamefont {Vartanyants}},\ }\href {\doibase
  10.1103/PhysRevE.88.044501} {\bibfield  {journal} {\bibinfo  {journal} {Phys.
  Rev. E}\ }\textbf {\bibinfo {volume} {88}},\ \bibinfo {pages} {044501}
  (\bibinfo {year} {2013}{\natexlab{b}})}\BibitemShut {NoStop}%
\bibitem [{\citenamefont {Zozulya}\ \emph {et~al.}(2012)\citenamefont
  {Zozulya}, \citenamefont {Bondarenko}, \citenamefont {Schavkan},
  \citenamefont {Westermeier}, \citenamefont {Gr\"{u}bel},\ and\ \citenamefont
  {Sprung}}]{Zozulya2012}%
  \BibitemOpen
  \bibfield  {author} {\bibinfo {author} {\bibfnamefont {A.~V.}\ \bibnamefont
  {Zozulya}}, \bibinfo {author} {\bibfnamefont {S.}~\bibnamefont {Bondarenko}},
  \bibinfo {author} {\bibfnamefont {A.}~\bibnamefont {Schavkan}}, \bibinfo
  {author} {\bibfnamefont {F.}~\bibnamefont {Westermeier}}, \bibinfo {author}
  {\bibfnamefont {G.}~\bibnamefont {Gr\"{u}bel}}, \ and\ \bibinfo {author}
  {\bibfnamefont {M.}~\bibnamefont {Sprung}},\ }\href {\doibase
  10.1364/OE.20.018967} {\bibfield  {journal} {\bibinfo  {journal} {Opt.
  Express}\ }\textbf {\bibinfo {volume} {20}},\ \bibinfo {pages} {18967}
  (\bibinfo {year} {2012})}\BibitemShut {NoStop}%
\bibitem [{\citenamefont {Huang}\ \emph {et~al.}(1989)\citenamefont {Huang},
  \citenamefont {Nounesis}, \citenamefont {Geer}, \citenamefont {Goodby},\ and\
  \citenamefont {Guillon}}]{Huang1989}%
  \BibitemOpen
  \bibfield  {author} {\bibinfo {author} {\bibfnamefont {C.~C.}\ \bibnamefont
  {Huang}}, \bibinfo {author} {\bibfnamefont {G.}~\bibnamefont {Nounesis}},
  \bibinfo {author} {\bibfnamefont {R.}~\bibnamefont {Geer}}, \bibinfo {author}
  {\bibfnamefont {J.~W.}\ \bibnamefont {Goodby}}, \ and\ \bibinfo {author}
  {\bibfnamefont {D.}~\bibnamefont {Guillon}},\ }\href {\doibase
  10.1103/PhysRevA.39.3741} {\bibfield  {journal} {\bibinfo  {journal} {Phys.
  Rev. A}\ }\textbf {\bibinfo {volume} {39}},\ \bibinfo {pages} {3741}
  (\bibinfo {year} {1989})}\BibitemShut {NoStop}%
\bibitem [{\citenamefont {Aeppli}\ and\ \citenamefont
  {Bruinsma}(1984)}]{Aeppli1984}%
  \BibitemOpen
  \bibfield  {author} {\bibinfo {author} {\bibfnamefont {G.}~\bibnamefont
  {Aeppli}}\ and\ \bibinfo {author} {\bibfnamefont {R.}~\bibnamefont
  {Bruinsma}},\ }\href {\doibase 10.1103/PhysRevLett.53.2133} {\bibfield
  {journal} {\bibinfo  {journal} {Phys. Rev. Lett.}\ }\textbf {\bibinfo
  {volume} {53}},\ \bibinfo {pages} {2133} (\bibinfo {year}
  {1984})}\BibitemShut {NoStop}%
\bibitem [{\citenamefont {Davey}\ \emph {et~al.}(1984)\citenamefont {Davey},
  \citenamefont {Budai}, \citenamefont {Goodby}, \citenamefont {Pindak},\ and\
  \citenamefont {Moncton}}]{Davey1984}%
  \BibitemOpen
  \bibfield  {author} {\bibinfo {author} {\bibfnamefont {S.~C.}\ \bibnamefont
  {Davey}}, \bibinfo {author} {\bibfnamefont {J.}~\bibnamefont {Budai}},
  \bibinfo {author} {\bibfnamefont {J.~W.}\ \bibnamefont {Goodby}}, \bibinfo
  {author} {\bibfnamefont {R.}~\bibnamefont {Pindak}}, \ and\ \bibinfo {author}
  {\bibfnamefont {D.~E.}\ \bibnamefont {Moncton}},\ }\href {\doibase
  10.1103/PhysRevLett.53.2129} {\bibfield  {journal} {\bibinfo  {journal}
  {Phys. Rev. Lett.}\ }\textbf {\bibinfo {volume} {53}},\ \bibinfo {pages}
  {2129} (\bibinfo {year} {1984})}\BibitemShut {NoStop}%
\bibitem [{\citenamefont {Clark}\ \emph {et~al.}(1983)\citenamefont {Clark},
  \citenamefont {Ackerson},\ and\ \citenamefont {Hurd}}]{Clark1983}%
  \BibitemOpen
  \bibfield  {author} {\bibinfo {author} {\bibfnamefont {N.~A.}\ \bibnamefont
  {Clark}}, \bibinfo {author} {\bibfnamefont {B.~J.}\ \bibnamefont {Ackerson}},
  \ and\ \bibinfo {author} {\bibfnamefont {A.~J.}\ \bibnamefont {Hurd}},\
  }\href {\doibase 10.1103/PhysRevLett.50.1459} {\bibfield  {journal} {\bibinfo
   {journal} {Phys. Rev. Lett.}\ }\textbf {\bibinfo {volume} {50}},\ \bibinfo
  {pages} {1459} (\bibinfo {year} {1983})}\BibitemShut {NoStop}%
\bibitem [{Har()}]{Harmonics_q}%
  \BibitemOpen
  \href@noop {} {}\bibinfo {note} {We should note here that the general
  expression for the $q-$dependence of the harmonics $I_n(q)$ is currently
  unknown and requires further theoretical development.}\BibitemShut {Stop}%
\bibitem [{NOT()}]{NOTE1}%
  \BibitemOpen
  \href@noop {} {}\bibinfo {note} {We note that the values of $\lambda(T)$ and
  $\mu(T)$ are similar to the values determined in Fig.
  \ref{Fig4}(c).}\BibitemShut {Stop}%
\bibitem [{\citenamefont {Bevington}\ and\ \citenamefont
  {Keith}(2003)}]{Bevington2003}%
  \BibitemOpen
  \bibfield  {author} {\bibinfo {author} {\bibfnamefont {P.~R.}\ \bibnamefont
  {Bevington}}\ and\ \bibinfo {author} {\bibfnamefont {R.~D.}\ \bibnamefont
  {Keith}},\ }\href@noop {} {\emph {\bibinfo {title} {Data Reduction and Error
  Analysis for the Physical Sciences}}}\ (\bibinfo  {publisher} {McGraw-Hill},\
  \bibinfo {year} {2003})\BibitemShut {NoStop}%
\end{thebibliography}%
	
\clearpage

	%\usepackage{setspace}
	%\makeatletter
	\setcounter{page}{1} 
	\setcounter{figure}{0} 
	\setcounter{equation}{0}
	\setcounter{enumiv}{0} 
	\renewcommand{\theequation}{S\arabic{equation}}
	\renewcommand{\thefigure}{S\arabic{figure}}
	\renewcommand{\bibnumfmt}[1]{[S#1]}
	\renewcommand{\citenumfont}[1]{S#1}
	%\captionsetup[figure]{labelformat=simple, labelsep=period}

		\begin{center}
			\textbf{\LARGE Supplemental Materials}
		\end{center}	
		\section{Positional correlation length}
\begin{figure}[b]
			\includegraphics[width=0.5\linewidth]{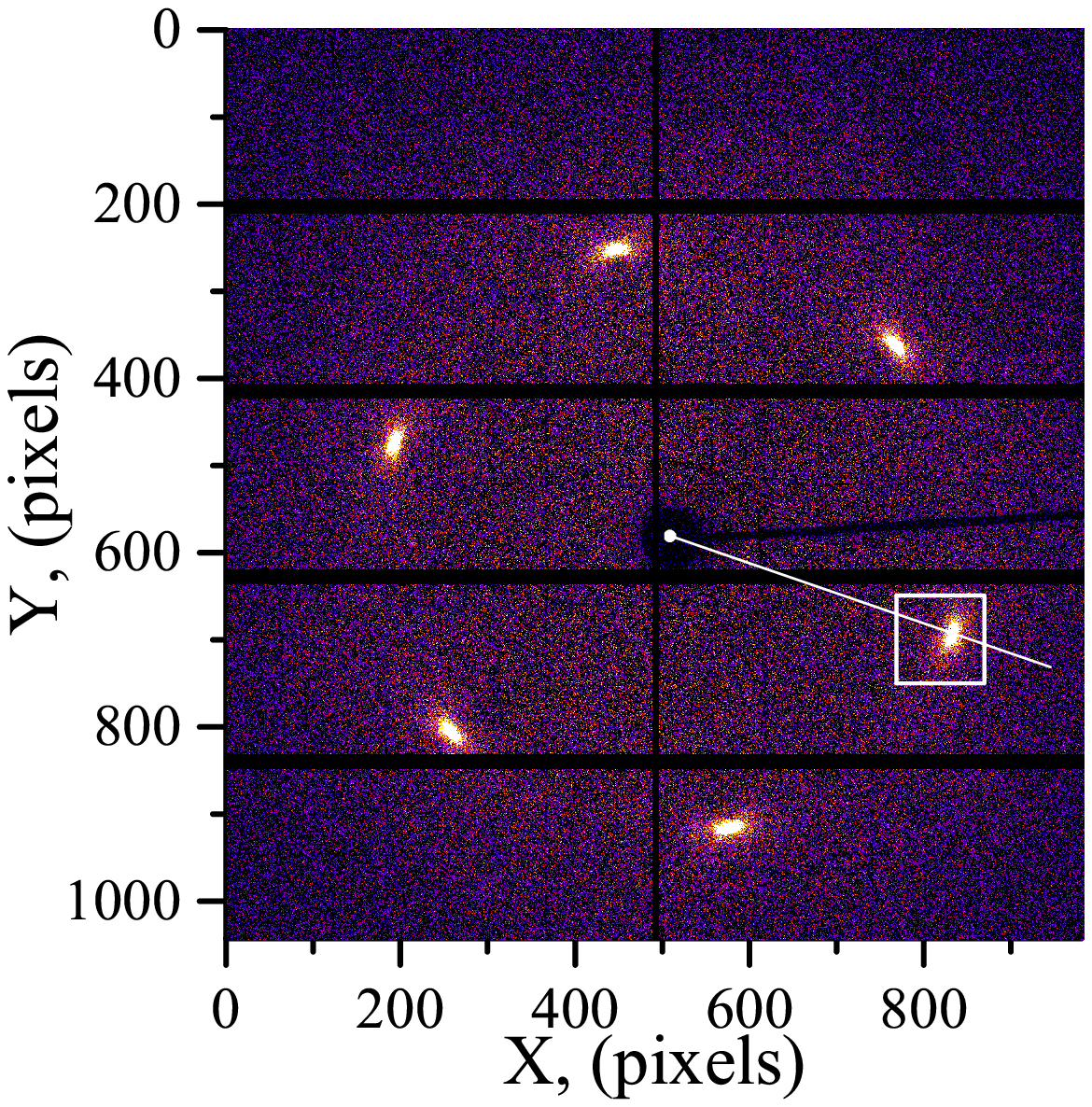}
			\caption{The single diffraction pattern at the temperature $T = 64.0~\degree$C. The area for the searching of the peaks center position is shown by a rectangle, and the line scan through the center of the peak is shown by a straight line.}
			\label{FigS1}
		\end{figure}

		The positional correlation length was determined by the analysis of the radial intensity scan through the diffraction peak. 
		For each diffraction pattern the area of $100\times100$ pixels around the diffraction peak (within the range $X\in[770, 870]$ and $Y\in[650, 750]$) was considered [Fig. \ref{FigS1}]. 
		The intensity $I^{j}(q)$ for $j$-th diffraction pattern  was measured along the line connecting the center of the pattern $(X_C=509.9,~Y_C=583.4)$ and the pixel with maximum intensity in this area, which corresponds to the center of the diffraction peak. 
		For the calculation of the intensity between the detector pixels bilinear interpolation was used. 
		Then all obtained profiles $I^j(q)$ at each temperature point were averaged over $N=100$ diffraction patterns
		\begin{equation}
		\label{EqS1}
		I_{tot}(q)=\frac{1}{N}\sum_{j=1}^{N}I^{j}(q).
		\end{equation}

\begin{figure}
			\includegraphics[width=1\linewidth]{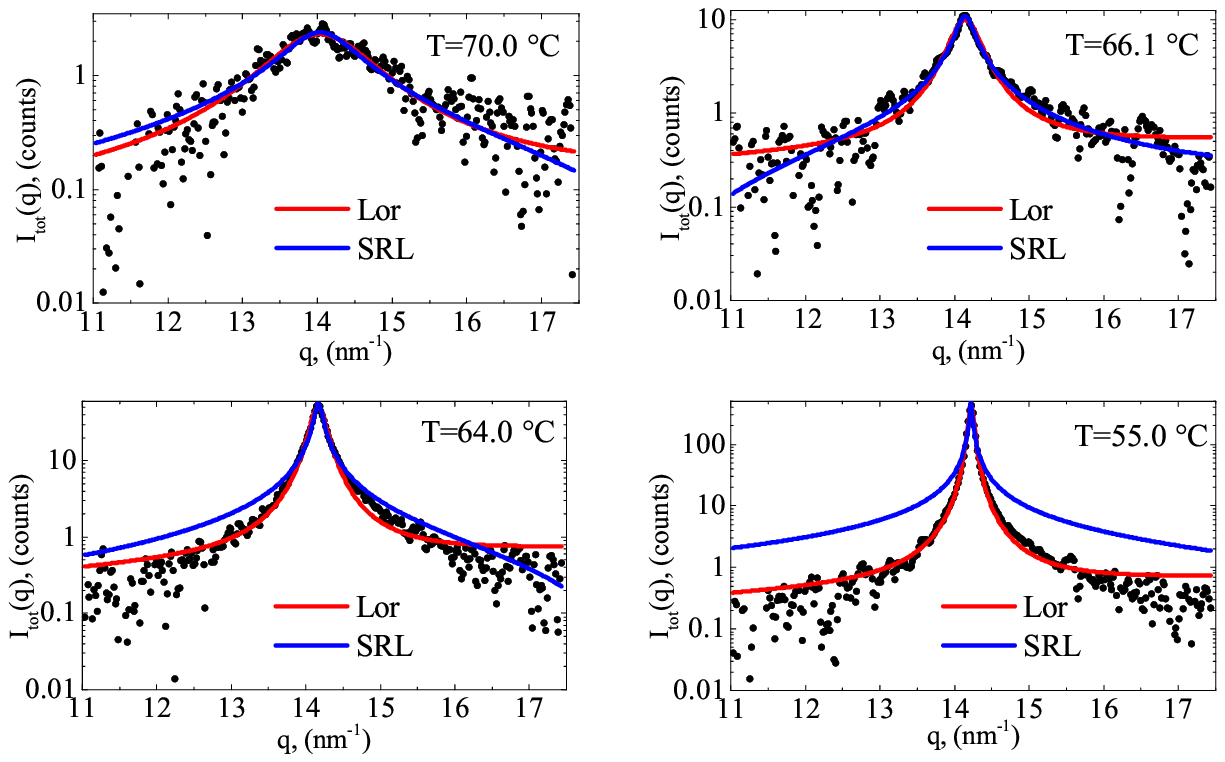}
			\caption{The fitting of $I_{tot}(q)$ with Lorentzian (red) and SRL (blue) functions for the temperatures $T = 70.0~\degree$C, $T = 66.1~\degree$C, $T = 64.0~\degree$C and $T = 55.0~\degree$C.}
			\label{FigS2}
		\end{figure}

		Usually in the systems with short-range positional order the diffraction peak $I_{tot}(q)$ can be described by Lorentzian function. However according to the phenomenological theory of Aeppli and Bruinsma $[\text{S}1]$  the shape of the radial scan through the center of the diffraction spot should have a form of a square root Lorentzian in the vicinity of the hexatic-smectic phase transition. In this work the averaged diffraction peaks $I_{tot}(q)$ were fitted whithin the range $q\in[11.0\:\text{nm}^{-1},\:17.5\:\text{nm}^{-1}]$ with both Lorentzian (Lor) and square-root Lorentzian (SRL) functions at each temperature point [Fig.~\ref{FigS2}]:
		\begin{equation}
		\label{Lor}
		I_{tot}^{Lor}(q)=A+B\cdot q + C \frac{1}{(q-q_0)^2+\gamma^2},
		\end{equation}
		\begin{equation}
		\label{SRL}
		I_{tot}^{SRL}(q)=A+B\cdot q + C \sqrt{\frac{1}{(q-q_0)^2+\gamma^2}}.
		\end{equation}

\begin{figure}
			\includegraphics[width=1\linewidth]{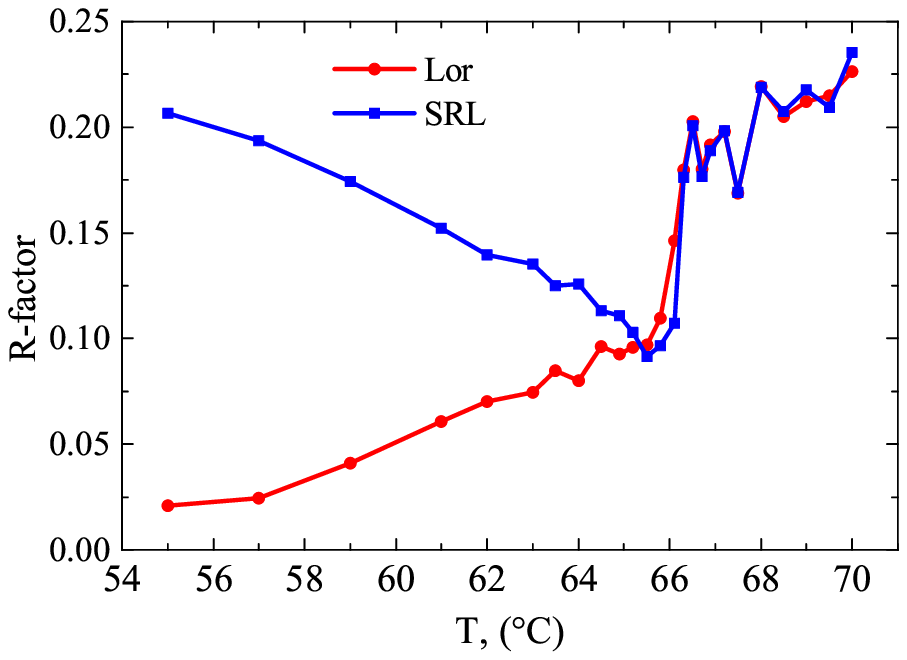}
			\caption{Temperature dependence of R-factor for the fitting of the experimental data $I_{tot}(q)$ with Lorentzian (red) and SRL (blue) functions.}
			\label{FigS3}
		\end{figure}

		The quality of the fitting can be estimated by considering the so-called R-factor, which directly corresponds to the residual sum of squares:
		\begin{equation}
		\label{RSS}
		R=\sqrt{\frac{\sum_{i=1}^{i_{max}}\big(I_{tot}^{exp}(q_i)-I_{tot}^{mod}(q_i)\big)^2}{\sum_{i=1}^{i_{max}}\big(I_{tot}^{exp}(q_i)\big)^2}}.
		\end{equation}
		In case of ideal fitting R-factor equals to zero and its value is larger if the experimental data do not match precisely with the fitting function. Smaller value of R-factor indicates better approximation of experimental data with the fitting function.  The comparison of two fittings with Lorentzian and SRL functions are shown in Fig.~\ref{FigS2}, and the values of R-factor for all temperature points are shown in Fig.~\ref{FigS3}.
		R-factor criterion reveals that the fitting with Lorentzian function is better for all temperatures except the temperature region $65.5~\degree$C$\leq T\leq 66.1~\degree$C of the hexatic-smectic phase transition, that is in a good agreement with theory $[\text{S}1]$.
		
\begin{figure}
			\includegraphics[width=1\linewidth]{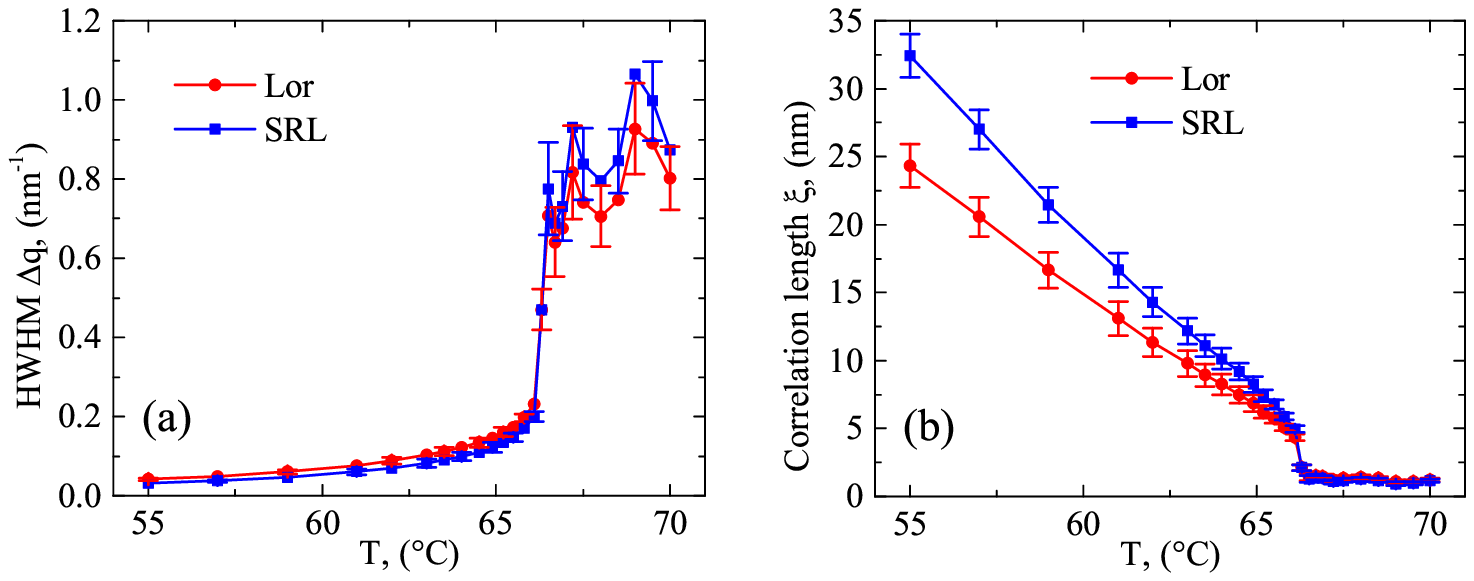}
			\caption{(a) The temperature dependence of the HWHM $\Delta q$ for fitting with Lorentzian and SRL functions. (b) The temperature dependence of the correlation length $\xi$ for fitting functions with Lorentzian and SRL functions.}
			\label{FigS4}
		\end{figure}

		The half width at half maxima (HWHM) of the peak was calculated as $\Delta q=\gamma$ (for Lorentzian function) and $\Delta q=\gamma \sqrt{3}$  (for SRL function). 
		The temperature dependence of the HWHMs $\Delta q$ is shown in Fig.~\ref{FigS4}(a) for both Lorentzian and SQR fitting. 
		The HWHM decreases upon cooling below the phase transition temperature, that indicates the coupling between positional and bond-orientational (BO) order in the hexatic phase.   
		
		We defined the positional correlation length in a conventional manner as $\xi=1/\Delta q$ $[\text{S}2]$.
		In Fig. \ref{FigS4}(b) the temperature dependence of $\xi$ is shown for both Lorentzian and SRL fittings.
		Since the values of the correlation length $\xi$ obtained by the fitting with Lorentzian and SRL functions are almost the same within the temperature region $65.5~\degree$C$\,\leq\,T\,\leq\,66.1~\degree$C, in our work we used the values of $\xi$, which were obtained by the fitting of the experimental data with Lorentzian function for all temperature points. 
		
		\section{Determination of a hexatic domain orientation}
\begin{figure}
			\includegraphics[width=1\linewidth]{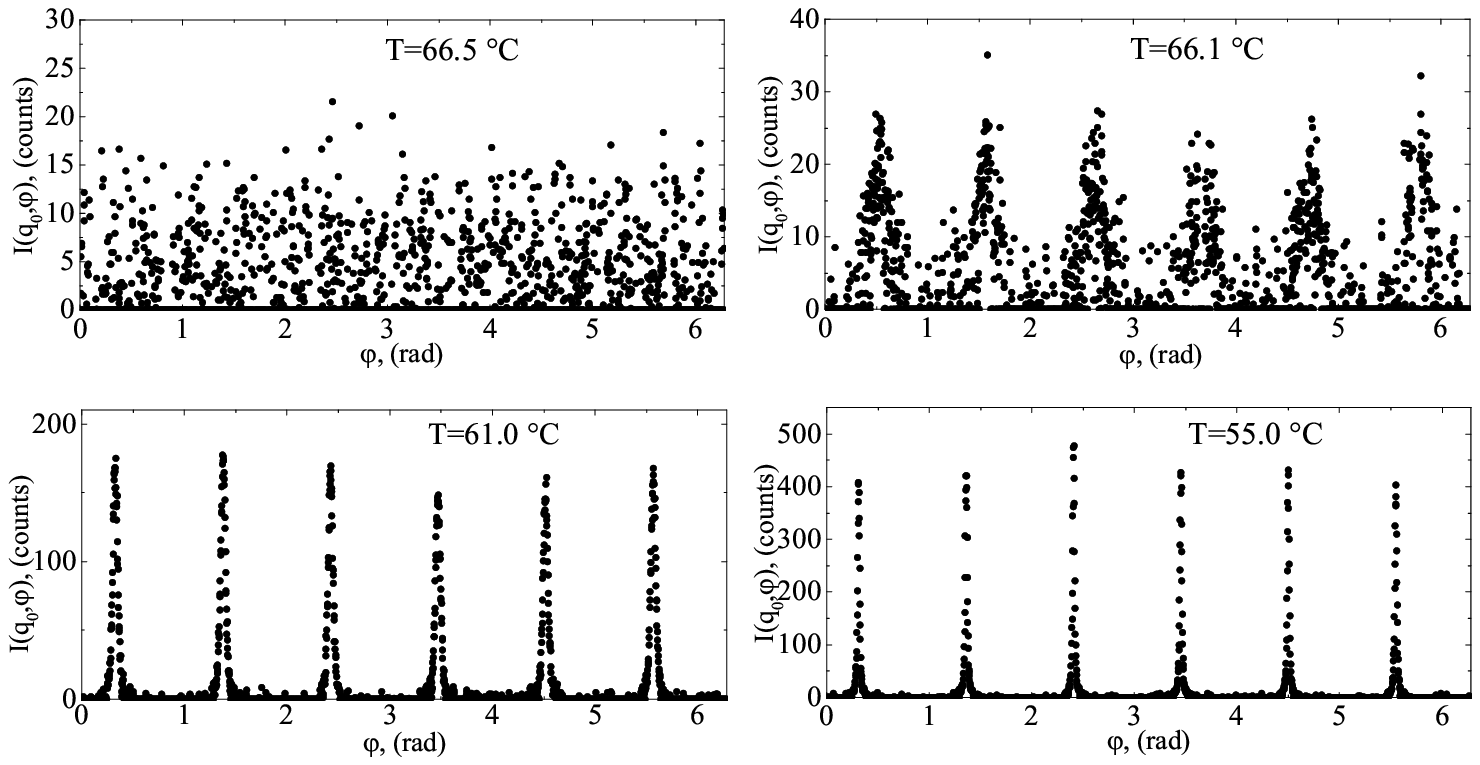}
			\caption{The scattered intensity $I(q_0,\phi)$ at the temperatures $T~=66.5~\degree$C, $T~=66.1~\degree$C, $T~=61.0~\degree$C and $T~=55.0~\degree$C.}
			\label{FigS5}
		\end{figure}

		The scattered intensity for $j$-th diffraction pattern [Fig. \ref{FigS5}] was decomposed into a cosine Fourier series:
		\begin{equation}
		\label{Fourier_decomposition}
		I^{j}(q,\phi)=I_0^{j}(q)+2\sum_{n=1}^{+\infty}|I_{n}^{j}(q)|\cos \big(n\phi-\phi_n^{j}(q)\big),
		\end{equation}
		\begin{equation}
		\label{Fourier_decomposition_2}
		|I_{n}^{j}(q)|\exp{\big(i\phi_n^{j}(q)\big)}=\frac{1}{2\pi}\int\limits_{0}^{2\pi}I^{j}(q,\phi)\exp{(-i n \phi)}\,\mathrm{d}\phi,
		\end{equation}
		where $(q, \phi)$ are the polar coordinates in the detector plane, $I_0^{j}(q)$ is the scattered intensity averaged over a scattering ring of a radius $q$, and $|I_n^{j}(q)|$ and $\phi_n^{j}(q)$ are the magnitude and phase of the $n$-th Fourier component (FC).
		We used bilinear interpolation to calculate the intensity $I(q_0,\phi)$ at the points between the pixels.
		To determinate the intermolecular bond orientations in the hexatic phase we considered the dominant FC of the order $n=6$, calculated at the value $q=q_0$.  
		Since the term $|I_{6}^{j}(q)|\cos \big(6\phi-\phi_6^{j}(q)\big)$ in Eq. (\ref{Fourier_decomposition}) is a periodic function of an angle $\phi$ with a period of $\frac{2\pi}{6}$, the values of the phase $\phi^{j}_6(q)$ can be chosen uniquely to be within the range $\phi^{j}_6(q_0)\in[\phi_{6}(q_0)-\frac{\pi}{6}, \phi_{6}(q_0)+\frac{\pi}{6}]$.
		The value $\phi_{6}(q_0)$ was chosen in such way, that the angular position of the center of one diffraction peak in the hexatic phase (specified by an arrow in Fig. 1(b) in the main text) was always within the range $[\phi_{6}(q_0)-\frac{\pi}{6}, \phi_{6}(q_0)+\frac{\pi}{6}]$.
		
		\section{Bond-orientational order parameters $C_{6m}$}
		Currently the general expression for the hexatic structure factor, depending on harmonic number and the temperature variance from the transition point, is unknown. 
		For small $m$ the $q$-dependence of the FCs $I_{6m}(q)$ can be well fitted by the SRL function. 
		However, for larger $m$ the wings of the curves becomes steeper, and this approximation fails to describe the experimental dependence $I_{6m}(q)$. In present work we fitted the first three components $I_{6m}(q)$ ($m=1,2,3$) with SRL function and the higher components were fitted by Lorentzian function.
		
		In order to calculate the bond-orientational (BO) order parameters $C_{6m}=(I_{6m} (q_0))/(I_0 (q_0))$ we considered independently the $q$-dependence of the Fourier coefficients of intensity $I_{6m}(q)$ and $I_{0}(q)$ obtained by means of angular x-ray cross-correlation analysis (XCCA). Since $I_{6m}(q)$ and $I_{0}(q)$ have well-determined maxima at the same position $q=q_0$, we calculated the values $C_{6m}$ as a ratio of the maxima value of $I_{6m}(q)$ to the maxima value of $I_{0}(q)$:
		\begin{equation}
		\label{Coeff}
		C_{6m}=\frac{\max{I_{6m}(q)}}{\max{I_{0}(q)}},
		\end{equation}
		where $q$ is within the range $q\in[13\:\text{nm}^{-1},\:15\:\text{nm}^{-1}]$.
		\section{Multicritical scaling theory}
\begin{figure} 
			\includegraphics[width=1.0\linewidth]{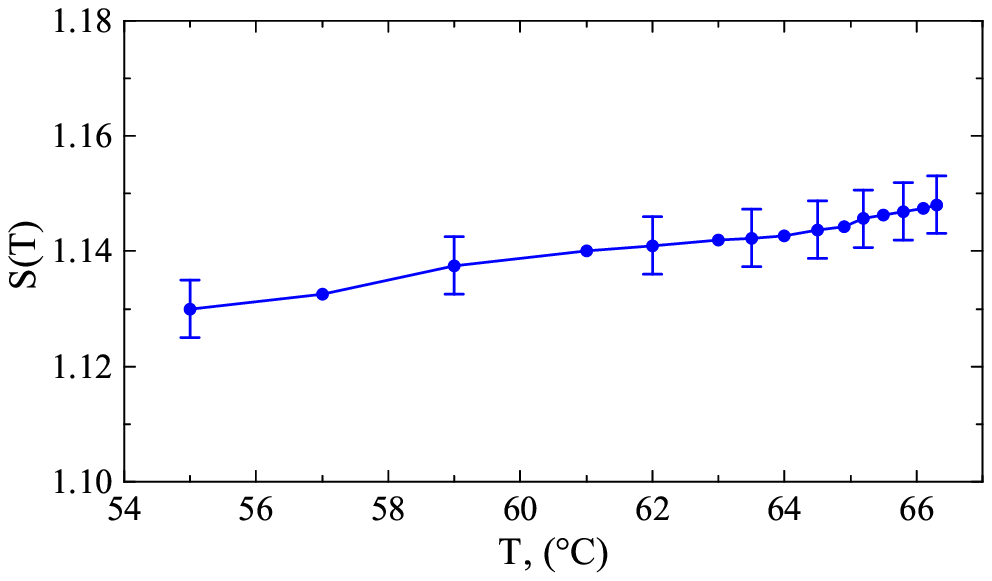}
			\caption{Temperature dependence of the correction coefficient $S(T)$.}
			\label{FigS6}
		\end{figure}

\begin{figure}
			\includegraphics[width=1\linewidth]{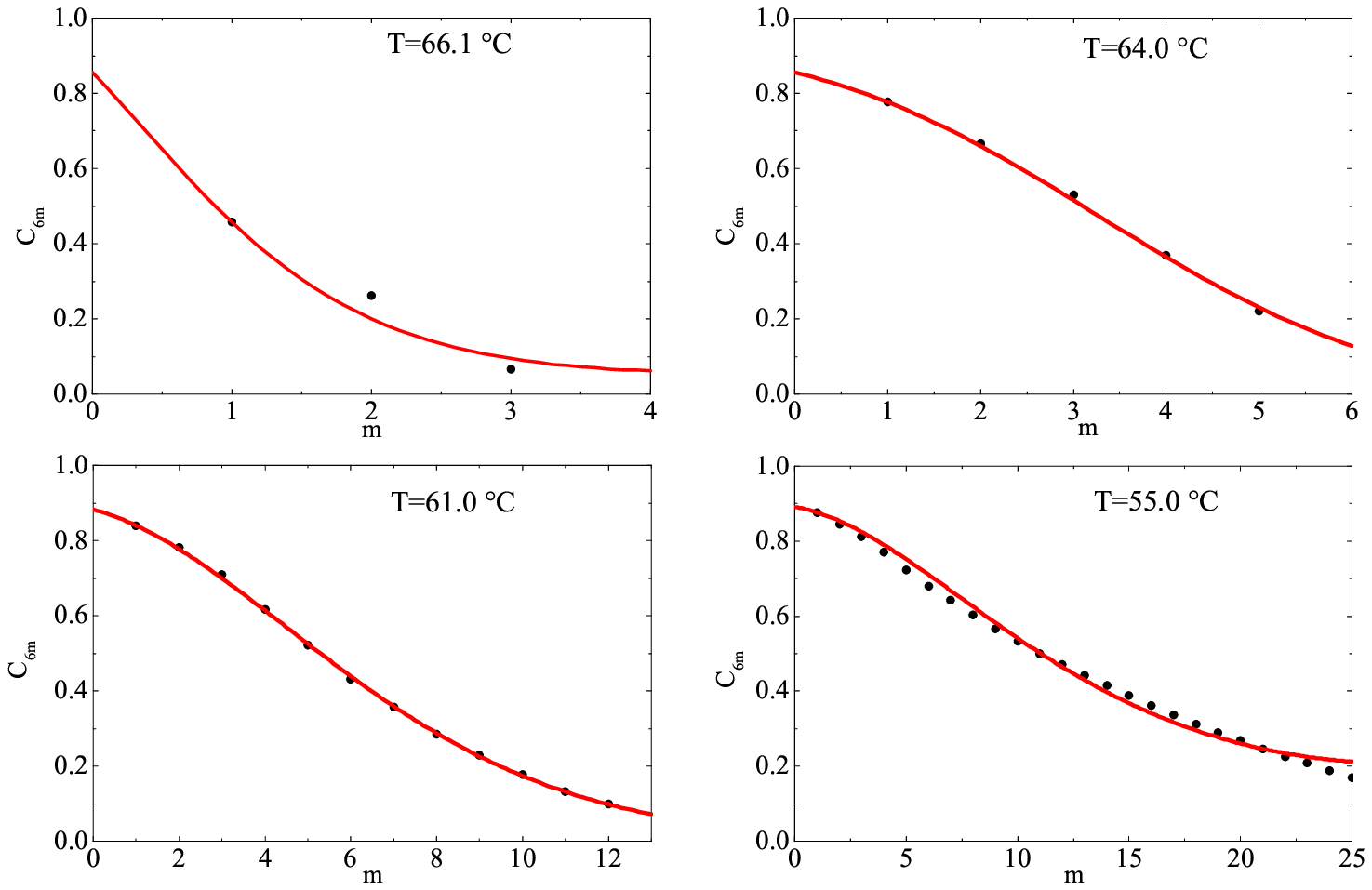}
			\caption{The fitting of $C_{6m}$ with function \ref{Mult2} at the temperatures $T = 66.1~\degree$C, $T = 64.0~\degree$C, $T = 61.0~\degree$C and $T = 55.0~\degree$C.}
			\label{FigS7}
		\end{figure}

		According to the multicritical scaling theory (MCST) the values $C_{6m}$ obey the scaling law $C_{6m}=(C_6 )^{\sigma_m}$, where $\sigma_m=m+x_m\cdot m \cdot (m-1)$ and $x^{(1)}_m=\lambda (T)-\mu (T)m$ $[\text{S}3,\,\text{S}4]$. 
		In order to compensate the contribution of the noise in $I_0(q_0)$ and consequently in $C_{6m}$ [see Eq. (\ref{Coeff})] we introduced the correction coefficient $S(T)$ into the scaling relation:
		\begin{equation}
		\label{Mult1}
		S(T)C_{6m}=(S(T)C_6)^{\sigma_m}.
		\end{equation}
		At each temperature point the experimentally obtained values $C_{6m} \quad (m=1,2,3,\dots)$ were fitted with the following function
		\begin{equation}
		\label{Mult2}
		C_{6m}=\frac{1}{S(T)}(S(T)C_6)^{\sigma_m}
		\end{equation}
		with three free parameters $S(T)$, $\lambda(T)$ and $\mu(T)$. The temperature dependence of $S(T)$ is shown in Fig.~\ref{FigS6}. It slightly decreases upon cooling down from 1.15 at $T = 66.1~\degree$C to 1.13 at $T = 55.0~\degree$C. The examples of the fitting of $C_{6m}$ with Eq. (\ref{Mult2}) is shown in Fig.~\ref{FigS7}.

\begin{figure}[H]
			\includegraphics[width=1\linewidth]{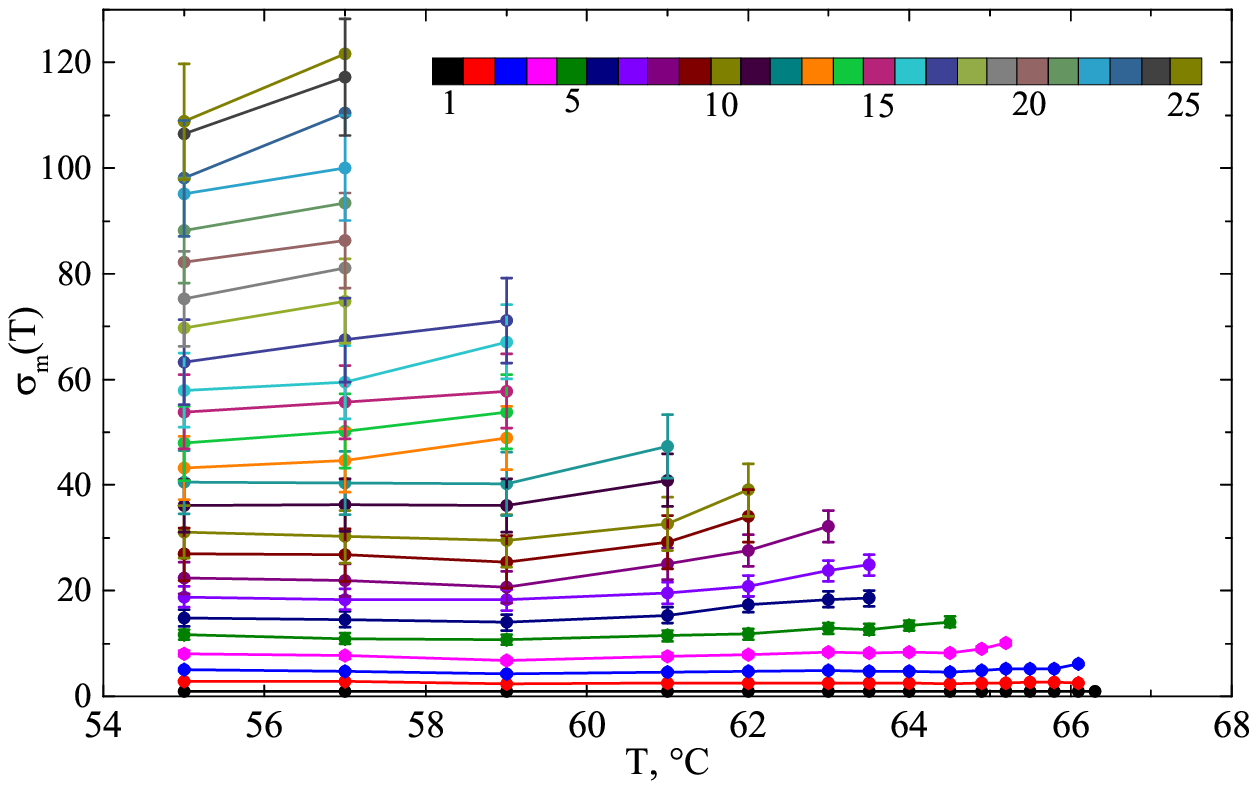}
			\caption{Temperature dependence of exponents $\sigma_m$ for all harmonics.}
			\label{FigS8}
		\end{figure}

		With known values of the correction coefficient $S(T)$ one can directly calculate values of $\sigma_m(T)$ at each temperature point according to the formula:
		\begin{equation}
		\label{Mult3}
		\sigma_{m}(T)=\frac{\ln{\big(SC_{6m}\big)}}{\ln{\big(SC_{6}\big)}}.
		\end{equation}
		The exponents $\sigma_m(T)$ are shown in Fig. \ref{FigS8} for all temperatures.
		One can clearly see, that the values of $\sigma_m(T)$ are practically independent of temperature for lower harmonics and change within the error for higher harmonics. This allows us to calculate the temperature-averaged values $\langle \sigma_m \rangle _T$. 

%\begin{thebibliography}{9}% Bibliography
%	\bibitem{Aeppli1984Supp}
%	G. Aeppli and R. Bruinsma, Phys. Rev. Lett. \textbf{53}, 2133 (1984)
%	\bibitem{Aharony1986Supp}
%	A. Aharony, R. J. Birgeneau, J. D. Brock, and J. D. Litster, Phys. Rev. Lett. \textbf{57}, 1012 (1986)
%	\bibitem{Brock1986Supp}
%	J. D. Brock, A. Aharony, R. J. Birgeneau, K. W. Evans-Lutterodt, J. D. Litster, P. M. Horn, G. B. Stephenson, and A. R. Tajbakhsh, Phys. Rev. Lett. \textbf{57}, 98 (1986)
% %	W. H. de Jeu, B. I. Ostrovskii, and A. N. Shalaginov, Rev. Mod. Phys. \textbf{75}, 181 (2003)	
%\end{thebibliography}

\section{References}
	$[\text{S}1]$ G. Aeppli and R. Bruinsma, Phys. Rev. Lett. \textbf{53}, 2133 (1984)
	
	$[\text{S}2]$ W. H. de Jeu, B. I. Ostrovskii, and A. N. Shalaginov, Rev. Mod. Phys. \textbf{75}, 181 (2003)	
	
	$[\text{S}3]$ A. Aharony, R. J. Birgeneau, J. D. Brock, and J. D. Litster, Phys. Rev. Lett. \textbf{57}, 1012 (1986)
	
	$[\text{S}4]$ J. D. Brock, A. Aharony, R. J. Birgeneau, K. W. Evans-Lutterodt, J. D. Litster, P. M. Horn, G. B. Stephenson, and A. R. Tajbakhsh, Phys. Rev. Lett. \textbf{57}, 98 (1986)

\end{document}